%% file: TI.tex
\documentclass[preprint,3p,sort&compress]{elsarticle}

\usepackage{layouts}
\usepackage{lineno,hyperref}
\usepackage{amsmath}
\usepackage{amssymb}
\usepackage{dsfont}
\usepackage{nicefrac}
\usepackage{here}
\usepackage{multirow}
\usepackage{multicol}
\usepackage{diagbox}
\usepackage{amssymb}
\usepackage{caption}
\usepackage{subcaption}
\usepackage{color}
\usepackage{titlesec}
\usepackage{appendix}
\usepackage{stmaryrd}
\usepackage{booktabs} 
\usepackage{placeins}
\usepackage{soul}
\usepackage[version=3]{mhchem} 
\usepackage{mathtools} 
\usepackage[dvipsnames, svgnames]{xcolor}
\usepackage{siunitx}		
\usepackage[inline]{trackchanges}
\usepackage{bm}
\usepackage{cancel}
\usepackage{tabularx}
\usepackage[export]{adjustbox}
\usepackage{cleveref}
\usepackage{MnSymbol}
\usepackage{ulem}

\definecolor{green_py}{rgb}{0.2, 0.4, 0}

\addeditor{R1}
\addeditor{R2}
\addeditor{R3}

\usepackage{algorithm}
\usepackage{algpseudocode}
\usepackage{tikz}		
\definecolor{green_pyt}{rgb}{0.192, 0.447, 0.282}
\definecolor{orange_pyt}{rgb}{0.65, 0.498, 0.188}
\definecolor{blue_pyt}{rgb}{0.403, 0.454, 0.976}
\definecolor{red_pyt}{rgb}{0.705, 0.031, 0.031}
\definecolor{white}{rgb}{1.0, 1.0, 1.0}

\tikzset{base/.style={draw, align=center, fill=orange_pyt, line width=0.4pt,minimum height=1ex,scale=1.45},
         test1/.style={base, diamond, aspect=1, text width=0.1em, inner sep=1.pt},
        }

\definecolor{bleuclair}{rgb}{0.266, 0.509, 0.972}
\definecolor{rosepale}{rgb}{0.937, 0.223, 0.223}
\definecolor{jaunepale}{rgb}{0.980, 0.890, 0.2}
\definecolor{vert}{rgb}{0.086, 0.694, 0.086}

\colorlet{rosepale}{rosepale!40}
\colorlet{bleuclair}{bleuclair!40}
\colorlet{jaunepale}{jaunepale!40}
\colorlet{vert}{vert!40}

\newcommand\Lfig{12cm} 
\newcommand\Hfig{7.40cm}

\modulolinenumbers[5]
\journal{Journal of Computational Physics}
\begin{document}
\begin{frontmatter}
\title{Extension of the Spectral Difference method to combustion}
\author[SAFRAN,CERFACS,ONERA]{Thomas Marchal\fnref{PhD}\corref{1}}
\ead{tmarchal@cerfacs.fr}
\author[ONERA]{Hugues Deniau\fnref{engineer}}
\ead{Hugues.Deniau@onera.fr}
\author[CERFACS]{Jean-François Boussuge\fnref{engineer}}
\ead{boussuge@cerfacs.fr}
\author[CERFACS]{Bénédicte Cuenot\fnref{engineer}}
\ead{cuenot@cerfacs.fr}
\author[SAFRAN]{Renaud Mercier\fnref{engineer}}
\ead{renaud-c.mercier@safrangroup.com}

\fntext[PhD]{Ph.D. Student}
\fntext[engineer]{Research engineer}
\cortext[1]{Corresponding author}

\address[SAFRAN]{Safran Tech, Digital Sciences and Technologies Department, Rue des Jeunes Bois, Châteaufort, Magny-Les-Hameaux 78114, France}
\address[CERFACS]{Centre Europ\'een de Recherche et de Formation Avanc\'ee en
Calcul Scientifique (CERFACS), \\ 42 avenue Gaspard Coriolis, 31057 Toulouse Cedex 01, France}
\address[ONERA]{ONERA/DMPE, Universit\'e de Toulouse, F-31055 Toulouse, France}

\begin{abstract}
A Spectral Difference (SD) algorithm on tensor-product elements which solves the reacting compressible Navier-Stokes equations (NSE) is presented. The classical SD algorithm is shown to be unstable when a multispecies gas where thermodynamic properties depend on temperature and species mass fractions is considered. In that case, a modification of the classical algorithm was successfully employed making it stable. It uses the fact that it is better for the multispecies case to compute primitive variables from conservative variables at solution points and then extrapolate them at flux points rather than extrapolating conservative variables at flux points and reconstruct primitive variables on these points. Characteristic, wall and symmetry boundary conditions for reactive flows in the SD framework are also introduced. They all use the polynomial form of the variables and of the fluxes to impose the correct boundary condition at a boundary flux point. Validation test cases on one-dimensional and two-dimensional laminar flames have been performed using both global chemistry and Analytically Reduced Chemistry (ARC). Results show excellent agreement with the reference combustion code AVBP validating the implementation of this SD method on laminar combustion. 
\end{abstract}


\end{frontmatter}

\input{1_intro}

\input{2_GoverningEquations}

\input{3_SpectralDifferenceMethod}
\input{5_NSCBCBoundaryConditions}

\input{4_WallBoundaryConditions}
\input{8_ValidationTestCases}
\input{Conclusion}


\newpage
\begin{appendices}
\input{Appendix_tangential_vectors}
\input{Appendix_system_to_solve_nscbc_inlet}

\input{Appendix_get_dQdU}
\end{appendices}

\clearpage
\bibliography{Biblio}

\end{document}

%% file: 1_intro.tex
\section{Introduction \label{sec:intro}}
Combustion phenomena are quite complex to understand and the use of numerical simulations, associated to experimental studies, is essential for the design of combustion engines. These simulations must be able to reproduce, as close as possible to reality, what happens inside a combustion chamber where there are multiple spatio-temporal scales, complex chemistry and flow structures but also strong variations of thermodynamic properties. During the last decades, large eddy simulations (LES) have appeared to be a good compromise between Reynolds Average Navier-Stokes (RANS) equations and direct numerical simulations (DNS) in the combustion field. However, LES require accurate spatial discretization with low dissipation and low dispersion properties.

With finite difference (FD) or finite volume (FV) techniques, increasing space accuracy means increasing the stencil of the scheme which by construction becomes very expensive when using unstructured meshes that are needed for complex geometries. On the contrary, recently developed high-order discontinuous methods give access to high-order accuracy on unstructured meshes with efficient parallelization. All these methods follow the same principle: they define a high-order representation (mostly a polynomial of degree $p$) of the variables within each mesh element and a high-order interpolation procedure. A major advantage is to exploit the possibility of increasing the spatial resolution by either increasing locally the polynomial order $p$, called $p$-refinement, or by doing local mesh refinement, called $h$-refinement. The first one reduces dissipation and dispersion errors in smooth flow regions whereas the second one is able to isolate regions with geometrical and physical discontinuities~\cite{chapelier2014evaluation,de2018use}.

The most famous high-order discontinuous approach is the discontinous Galerkin (DG) method developed by Reed and Hill in 1973 for the neutron transport equation~\cite{reed1973triangular}. Almost twenty years later, Cockburn and Shu applied the DG approach to conservation laws and more specifically to the Navier-Stokes equations (NSE)~\cite{cockburn1990runge,cockburn1997runge,cockburn2001runge}. Since then, it has been widely employed to perform LES simulations of multiple problems such as turbulent jets~\cite{anghan2019direct}, laminar to turbulent transition~\cite{beck2014high} or shock waves~\cite{renac2015aghora} and combustion applications~\cite{billet2014runge,lv2014discontinuous,lv2017high}. However, as the scheme order increases, the computational cost of the DG approach soars fastly. It is due to the fact that the DG method solves the weak integral form of the NSE so that integrals have to be evaluated using quadrature rules which becomes very costly as the scheme order rises~\cite{yu2014accuracy}.

Following this observation, new high-order discontinuous methods built to solve the strong form of the NSE have emerged and they are commonly splitted into two families. The first one is the Flux Reconstruction (FR) technique, also called the Correction Procedure for Reconstruction (CPR), developed by Huynh~\cite{huynh2007flux} in 2007. In this method the solution and the flux polynomials are collocated which implies that the flux divergence is no longer a polynomial of degree $p$ but also that the scheme is not conservative because the flux is discontinuous at element interfaces. Flux correction functions~\cite{allaneau2011connections} taken as polynomials of degree $p+1$ are then used to tackle these issues and make the FR method conservative. It has been applied to multiple configurations of flows over the past ten years~\cite{gao2009high,haga2011high,singh2019wall,haga2015large}. 

The second family of strong discontinuous approach, is the Spectral Difference (SD) method originally introduced by Kopriva and Colias in 1996~\cite{kopriva1996conservative,kopriva1996conservativeII} as the staggered-grid Chebyshev multidomain method. They applied it to structured quadrilateral elements using a tensor-product formulation. In 2006, Liu et al.~\cite{liu2006spectral} extended the method to triangular elements and conservation laws. Then, Wang et al.~\cite{wang2007spectral} used it for the Euler equations and the extension to NSE was done by May and Jameson~\cite{may2006spectral} for triangular grids and by Sun et al.~\cite{sun2007high} for hexahedral elements. For tensor-product cells such as quadrilaterals or hexahedrals, the standard SD method builts a polynomial of degree $p$ for the solution vector using solution values at what are called solution points (SP) and a polynomial of degree $p+1$ for the flux vector using flux values at another set of points called flux points (FP). It gives a scheme order of $p+1$. Recently, Chen et al.~\cite{chen2020collocated} presented a new formulation for this kind of elements where the flux divergence is built from the flux values at SP completed by flux values at interface FP. It avoids the need to interpolate from SP to internal FP which saves computational time. They called this approach the collocated-grid spectral difference (CGSD) method. Stability of the standard method on tensor-product elements was investigated by Van den Abeele et al.~\cite{van2008stability} and Jameson~\cite{jameson2010proof} who concluded that SP positions have no influence on stability whereas FP positions do. In particular, the SD method on tensor-product cells will be stable for all orders of accuracy if the interior FP are the roots of the corresponding Legendre polynomial of degree $p$. For triangular and tetrahedral cells, also called simplex cells, the stability also depends on FP but the set of stable FP is much harder to find. The use of Raviart-Thomas (RT) elements on triangles in SD, named as the SDRT method, was proven to be linearly stable up to the 4$^{th}$ order by May et al.~\cite{may2010analysis}. It seems to be the more promising approach and has been applied recently on various test cases~\cite{balan2012stable,li2019new,qiu2019high}. Moreover, Veilleux et al.~\cite{veilleux2021stable} extended the SDRT method on triangle elements up to the 6$^{th}$ order of accuracy and on tetrahedral elements up to the 3$^{rd}$ order~\cite{veilleux2021extension}. 

Some studies have tried to compare DG, FR and SD techniques but there is no final conclusion on which high-order method should be used for a given application. Actually, in Liang et al.~\cite{liang2013comparison}, it was shown that FR is most efficient and DG is the slowest but in terms of accuracy DG is the best and FR is the worst. For both efficiency and accuracy the SD method stands in between whereas in a more recent study, Cox et al.~\cite{cox2021accuracy} stated that the SD method is more robust and accurate than FR. This is very encouraging for pursuing the development of the SD method to simulate complex flows. The recent development of non-reflecting boundary conditions specially adapted to the SD and FR algorithms by Fievet et al~\cite{fievet2020strong,fievet2020numerical} on tensor-product cells, has opened a way to the simulation of combustion in confined media using these methods. They proved to be very promising for the field of turbulent combustion to capture the main flame structures and the most reacting species in a gas mixture. However, to the author's knowledge, only Gupta et al.~\cite{gupta2018numerical} have done combustion simulation using a strong discontinuous method, here the SD approach, and it was for one-dimensional detonation. Consequently, there is a need to explore the ability of such methods to compute combustion applications when a multispecies gas is considered along with combustion models and stiff reactive source terms. In this paper the SD code JAGUAR (proJect of an Aerodynamic solver using General Unstructured grids And high ordeR schemes)~\cite{cassagne2015high,hamri2015evaluation,veilleux2021stable,fievet2020strong,fievet2020numerical}, jointly developed by CERFACS and ONERA, is extended to reacting flows and validated on several academic test cases of increasing difficulty. The $p$-refinement recently implemented in JAGUAR and tested on the convection of an isentropic vortex and then on laminar flows over a cylinder~\cite{hirsch2021tilda} allowed a gain in reduction of the number of grid points from 38\% to more than 50\% compare to a uniform $p$ calculation to achieve the same level of error. The gain in computational time is then significant, between 25\% to 50\% compared to a uniform $p$ calculation. Similar observations were made when using the DG discretization on such academic test cases~\cite{naddei2019simulation}. Efficient $p$-refinement is very appealing for combustion applications where the flame zone requiring fine discretization is very localized. This motivates the present work, focusing on the resolution of combustion problems with the SD method and, as a first step, using a uniform polynomial degree $p$. It should be mentioned that this SD extension to combustion only considers here tensor-product elements since characteristic boundary conditions still need to be developed for simplex cells which is out of the scope of this paper. 

The paper is organised as follows. In Section~\ref{sec:GovEq}, the reacting NSE are recalled. Section~\ref{sec:SD_method} describes the SD discretization that is used on tensor-product elements. Sections~\ref{sec:charac_bc_SD} and ~\ref{sec:wall_bc_SD} explain how characteristic and wall boundary conditions are imposed within the SD framework for reactive flows. Section~\ref{sec:test_cases} shows results obtained with JAGUAR on one-dimensional and two-dimensional laminar flames. Finally, conclusions and perspectives are drawn in Section~\ref{sec:conclusion}.

%% file: 2_GoverningEquations.tex
\section{Governing Equations\label{sec:GovEq}}
\noindent In this paper, the three-dimensional reacting compressible NSE for multispecies gas with $N_{s}$ species are considered~\cite{poinsot2005theoretical}:
\begin{align}
    \frac{\partial \rho}{\partial t} + \nabla.\left(\rho \mathbf{u}\right) &= 0\label{eq:mass_cons_summary}\\
    \frac{\partial \rho\mathbf{u}}{\partial t} + \mathbf{\nabla}.\left(\rho \mathbf{u}\otimes\mathbf{u}\right) &= -\mathbf{\nabla} P + \mathbf{\nabla}.\left(\boldsymbol{\tau}\right) \label{eq:momentum_cons_summary}\\
    \frac{\partial \rho E}{\partial t} + \nabla.\left(\mathbf{u}\left(P+\rho E\right)\right) &
    = -\nabla.\left(\mathbf{q}\right) + \nabla.(\boldsymbol{\tau}.\mathbf{u})
     + \dot{\omega}_{T}\label{eq:energy_cons_summary} \\
    \frac{\partial \rho Y_{k}}{\partial t} + \nabla.\left(\rho\mathbf{u}Y_{k}\right) &= \nabla.\left(\mathbf{M}_{k}\right) +  \dot{\omega}_{k}\hspace{0.20 cm}\text{for}\hspace{0.20 cm} k=1,N_{s} \label{eq:species_cons_summary}
\end{align}
with $\rho$ the gas density, $\mathbf{u}=\left(u,v,w\right)^{\mathrm{T}}$ the velocity along physical coordinates $\left(x,y,z\right)$, $E$ the total energy per unit mass, $\mathbf{Y}=\left\{Y_{k}\right\}_{1\leq k\leq N_{s}}$ the vector of species mass fractions and $P$ the static pressure. In Eqs.~(\ref{eq:mass_cons_summary}-\ref{eq:species_cons_summary}) $\boldsymbol{\tau}$, $\mathbf{q}=\left(q_{x},q_{y},q_{z}\right)^{\mathrm{T}}$ and $\mathbf{M}_{k}=\left(M_{kx},M_{ky},M_{kz}\right)^{\mathrm{T}}$ are respectively the viscous stress tensor, the energy and species diffusion flux vectors defined as \cite{poinsot2005theoretical}:
\begin{equation}\begin{aligned}
    &\boldsymbol{\tau} = \mu\left[\nabla\mathbf{u}+\left(\nabla\mathbf{u}\right)^{\mathrm{T}}\right] - \frac{2}{3}\mu\nabla.\mathbf{u}\mathrm{I}
    \\
    &\mathbf{q} = -\lambda\mathbf{\nabla}T + \rho\sum\limits_{k=1}^{N_{s}}{h_{sk}Y_{k}\mathbf{V}_{k}}
    \\
    &\mathbf{M}_{k} = \rho\left( D_{k}\frac{W_{k}}{W}\mathbf{\nabla}X_{k} - Y_{k}\mathbf{V}^{c}\right)\hspace{0.20 cm}\text{for}\hspace{0.20 cm} k=1,N_{s}
    \label{eq:Def_energy_spec_fluxes}
\end{aligned}\end{equation}
where $\mathrm{I}$ is the identity matrix, $\lambda$ is the thermal conductivity, $\mu$ is the dynamic viscosity, $T$ is the temperature, $W$ is the molar mass of the mixture and $h_{sk}$, $X_{k}$ and $W_{k}$ are respectively the mass sensible enthalpy (tabulated every 100 K from JANAF thermochemical tables \cite{tables1971dr}), the mole fraction and the molar mass of species $k$. In Eq.~(\ref{eq:Def_energy_spec_fluxes}) $\mathbf{V}_{k}$ and $\mathbf{V}^{c}$ are the diffusion velocities and their associated correction velocity under Hirschfelder and Curtiss approximation \cite{poinsot2005theoretical}:
\begin{equation}\begin{aligned}
    &\mathbf{V}_{k} = -\frac{D_{k}W_{k}}{Y_{k}W}\mathbf{\nabla}X_{k}\hspace{0.20 cm}\text{for}\hspace{0.20 cm} k=1,N_{s} \\
    &\mathbf{V}^{c} = \sum\limits_{k=1}^{N_{s}}{D_{k}\frac{W_{k}}{W}\mathbf{\nabla}X_{k}}\hspace{0.20 cm}\text{for}\hspace{0.20 cm} k=1,N_{s}
\end{aligned}\end{equation}
with $D_{k}$ the diffusion coefficient of species $k$ into the rest of the mixture computed here assuming a constant Schmidt number $Sc_{k}$ for each species:
\begin{align}
    D_{k} = \frac{\mu}{\rho Sc_{k}}\hspace{0.20 cm}\text{for}\hspace{0.20 cm} k=1,N_{s}
    \label{eq:def_Dk_JAG_AVBP}
\end{align}
In practice, $\mathbf{M}_{k}$, $\mathbf{V}_{k}$ and $\mathbf{V}^{c}$ are computed using $\nabla Y_{k}$ instead of $\nabla X_{k}$ since $Y_{k}$ is directly solved through Eq.~(\ref{eq:species_cons_summary}). To do so, in their expressions the quantity $\left(W_{k}/W\right)\mathbf{\nabla}X_{k}$ is computed using:
\begin{equation}
    \frac{W_{k}}{W}\mathbf{\nabla}X_{k} = \frac{W_{k}}{W}\mathbf{\nabla}\left(\frac{Y_{k}W}{W_{k}}\right) = \frac{1}{W}\left(W\mathbf{\nabla}Y_{k}+Y_{k}\underbrace{\mathbf{\nabla}W}_{-W^{2}\mathbf{\nabla}\left(\frac{1}{W}\right)}\right) = \mathbf{\nabla}Y_{k}-Y_{k}W\mathbf{\nabla}\left(\frac{1}{W}\right)
    \label{eq:Lemme_nabla_Xk}
\end{equation}
where $\mathbf{\nabla}\left(1/W\right)$ is deduced from $\mathbf{\nabla}Y_{k}$ thanks to Eq.~(\ref{eq:Link_Grad_1W_Grad_Yk}):
\begin{equation}
   \mathbf{\nabla}\left(\frac{1}{W}\right) = \sum\limits_{k=1}^{N_{s}}{\frac{\mathbf{\nabla}Y_{k}}{W_{k}}}
   \label{eq:Link_Grad_1W_Grad_Yk}
\end{equation}
The dynamic viscosity is assumed to be independent of the gas composition and to be close to that of air. In this work, two different gas mixtures will be employed and they both use a power law for $\mu$ as a function of $T$:
\begin{align}
    \mu\left(T\right)=\mu_{ref}\left(\frac{T}{T_{ref}}\right)^{m}
\end{align}
where $T_{ref}$, $\mu_{ref}$ and $m$ are respectively a reference temperature, the dynamic viscosity at this reference temperature and the power law exponent. For the thermal conductivity, the Prandtl's number $Pr$ of the mixture is assumed to be constant so that $\lambda$ is obtained using Eq.~(\ref{eq:def_thermal_conduct}):
\begin{align}
    \lambda = \frac{\mu C_{p}}{Pr}
    \label{eq:def_thermal_conduct}
\end{align}
with $C_{p}$ the heat capacity at constant pressure of the mixture. The net production rate of each species $\dot{\omega}_{k}$ is computed using a classical Arrhenius's law \cite{poinsot2005theoretical} and the heat release $\dot{\omega}_{T}$ is deduced from:
\begin{equation}
    \dot{\omega}_{T} = - \sum\limits_{k=1}^{N_{s}}{\Delta h_{f,k}^{0}\dot{\omega}_{k}}
    \label{eq:Def_Heat_release}
\end{equation}
where $\Delta h_{f,k}^{0}$ is the formation enthalpy of species $k$ also obtained from JANAF thermochemical tables \cite{tables1971dr}.
Finally, the equations are closed assuming that the mixture and each species behave as an ideal gas where the static pressure is the sum of partial pressures of each species:
\begin{align}
P = \rho\frac{\overline{R}}{W}T = \rho\overline{R}T\sum\limits_{k=1}^{N_{s}}{\frac{Y_{k}}{W_{k}}}
\label{eq:perfect_gas_law}
\end{align}
with $\overline{R}=8.314\ \mathrm{J.mol^{-1}.K^{-1}}$ the ideal gas constant. Eqs.~(\ref{eq:mass_cons_summary}-\ref{eq:species_cons_summary}) can be recast into:
\begin{equation}
\frac{\partial \mathbf{U}}{\partial t}+\frac{\partial \mathbf{E}}{\partial x}+\frac{\partial \mathbf{F}}{\partial y}+\frac{\partial \mathbf{G}}{\partial z}=\mathbf{S}
\label{eq:NS_cons_phy}
\end{equation}
where $\mathbf{U}=\left(\rho,\rho u,\rho v,\rho w,\rho E,\rho Y_{1},\hdots,\rho Y_{N_{s}}\right)^{\mathrm{T}}$ is the vector of conservative variables, $\mathbf{E}=\mathbf{E}_{c}+\mathbf{E}_{d}$,  $\mathbf{F}=\mathbf{F}_{c}+\mathbf{F}_{d}$ and $\mathbf{G}=\mathbf{G}_{c}+\mathbf{G}_{d}$ are respectively the sum of convective and diffusive fluxes of $\mathbf{U}$ along $x$, $y$ and $z$ directions and $\mathbf{S}$ is a source term vector. They read as:
\begin{align}
\begin{array}{ccccccccc}
    &\mathbf{E}_{c} = (\rho u, &\rho u^{2} + P, & \rho v u, & \rho w u, & u\left(P+\rho E\right), & \rho u Y_{1}, &\hdots, & \rho u Y_{N_{s}})^{\mathrm{T}} \\
    &\mathbf{F}_{c} = (\rho v, &\rho u v, &\rho v^{2} + P, &\rho w v, & v\left(P+\rho E\right), & \rho v Y_{1}, &\hdots, & \rho v Y_{N_{s}})^{\mathrm{T}} \\
    &\mathbf{G}_{c} = (\rho w, &\rho u w, &\rho v w, &\rho w^{2} + P, &w\left(P+\rho E\right), & \rho w Y_{1}, &\hdots, & \rho w Y_{N_{s}})^{\mathrm{T}} \\
    &\mathbf{E}_{d} = (0, &-\tau_{11}, &-\tau_{21}, &-\tau_{21}, &\partial_{x}q_{x}-u\tau_{11}-v\tau_{12}-w\tau_{13}, &-\partial_{x}M_{1x}, &\hdots, & -\partial_{x}M_{N_{s}x})^{\mathrm{T}} \\
    &\mathbf{F}_{d} = (0, &-\tau_{12}, &-\tau_{22}, &-\tau_{32}, &\partial_{y}q_{y}-u\tau_{21}-v\tau_{22}-w\tau_{23}, &-\partial_{y}M_{1y}, &\hdots, & -\partial_{y}M_{N_{s}y})^{\mathrm{T}} \\
    &\mathbf{G}_{d} = (0, &-\tau_{13}, &-\tau_{23}, &-\tau_{33}, &\partial_{z}q_{z}-u\tau_{31}-v\tau_{32}-w\tau_{33}, &-\partial_{z}M_{1z}, &\hdots, & -\partial_{z}M_{N_{s}z})^{\mathrm{T}} \\
    &\mathbf{S} = (0, &0, &0, &0, &\dot{\omega}_{T}, &\dot{\omega}_{1}, &\hdots, & \dot{\omega}_{N_{s}})^{\mathrm{T}}
    \label{eq:def_fluxes_src_terms}
\end{array}
\end{align}

\noindent Actually, in Eq.~(\ref{eq:NS_cons_phy}) one equation is redundant since for a multispecies gas:
\begin{align}
    \rho = \sum\limits_{k=1}^{N_{s}}{\rho Y_{k}}
    \label{eq:get_rho_with_rhoYk}
\end{align}
Consequently, the mass conservation is still solved by the numerical scheme but the density is recomputed using Eq.~(\ref{eq:get_rho_with_rhoYk}) with the transported $\rho Y_{k}$ values.

%% file: 3_SpectralDifferenceMethod.tex
\section{The Spectral Difference method\label{sec:SD_method}}
\noindent In this section, the SD discretization process will be introduced for quadrilateral/hexahedral elements only. 

\subsection{Isoparametric transformation for hexahedral elements}
\noindent Let's consider a computational domain $\Omega$ divided into $N_{e}$ non-overlapping hexahedral elements inside which Eq.~(\ref{eq:NS_cons_phy}) is to be solved. Each element $\Omega_{e}$ of $\Omega$ will be transformed into a standard hexahedron $\mathcal{H}=\left\{\left(\xi, \eta,\zeta\right),\ 0\leq \xi, \eta,\zeta\leq 1\right\}$ following the transformation~\cite{sun2007high}:
\begin{align}
    \mathbf{x}\left(\boldsymbol{\xi}\right) = \sum\limits_{i=1}^{K}{M_{i}\left(\boldsymbol{\xi}\right)\mathbf{x}_{i}^{e}}
\end{align}
where $\mathbf{x}_{i}^{e}=\left(x_{i}^{e}, y_{i}^{e}, z_{i}^{e}\right)$ are the Cartesian coordinates in the physical domain of the $K$ vertices of $\Omega_{e}$ and $M_{i}\left(\boldsymbol{\xi}\right)$ are the shape functions. This transformation from the physical domain $\mathbf{x}=\left(x,y,z\right)$ to the reference domain $\boldsymbol{\xi}=\left(\xi,\eta,\zeta\right)$ is characterized by a Jacobian matrix $J$ along with its inverse (assuming a non-singular transformation) representing the reverse transformation:
\begin{equation}
    J = \left(
    \begin{array}{ccc}
        x_{\xi} & x_{\eta} & x_{\zeta} \\
         y_{\xi} & y_{\eta} & y_{\zeta} \\
          z_{\xi} & z_{\eta} & z_{\zeta} \\
    \end{array}
    \right)\hspace{0.25 cm}\text{and}\hspace{0.25 cm} J^{-1} = \left(
    \begin{array}{ccc}
        \xi_{x} & \xi_{y} & \xi_{z} \\
         \eta_{x} & \eta_{y} & \eta_{z} \\
          \zeta_{x} & \zeta_{y} & \zeta_{z} \\
    \end{array}
    \right)
\end{equation}
where the components of $J^{-1}$ are the grid metrics whose expressions can be found in~\cite{sun2007high}. Eq.~(\ref{eq:NS_cons_phy}) will be solved in the reference domain $\mathcal{H}$. Consequently, it will be written in this domain as \cite{sun2007high}:
\begin{equation}
    \frac{\partial \widehat{\mathbf{U}}}{\partial t}+\frac{\partial \widehat{\mathbf{E}}}{\partial \xi} + \frac{\partial \widehat{\mathbf{F}}}{\partial \eta} + \frac{\partial \widehat{\mathbf{G}}}{\partial \zeta} = \widehat{\mathbf{S}}
    \label{eq:NS_cons_iso_3rd}
\end{equation}
with
\begin{equation}
\left(\begin{array}{c}
     \widehat{\mathbf{E}}  \\
     \widehat{\mathbf{F}} \\
     \widehat{\mathbf{G}}
\end{array}
\right) = |J|J^{-1}.\left(\begin{array}{c}
     \mathbf{E}  \\
     \mathbf{F} \\
     \mathbf{G}
\end{array}
\right)
\label{eq:def_E_F_G_hat}
\end{equation}
where $\widehat{\mathbf{U}}=|J|\mathbf{U}$ and $\widehat{\mathbf{S}}=|J|\mathbf{S}$, $|J|$ being the determinant of $J$.

\subsection{General principle of the Spectral Difference discretization for a hyperbolic 1D-equation}
\label{sub:GenePrin1DSD}
\noindent The SD discretization will be described for an order of accuracy of $p+1$ inside each element $\Omega_{e}$ to define some notations and give some context for the parts where boundary conditions will be explained. As this work focuses only on tensor-product elements such as quadrilaterals or hexahedrals, the discretization process can be presented on a one-dimensional domain. Each element $\Omega_{e}$ is then a 1D line $\left[a,b\right]$ that will be transformed into a standard $\left[0,1\right]$ line $\mathcal{L}$ where Eq.~(\ref{eq:NS_cons_iso_1D}) has to be solved:
\begin{align}
    \frac{\partial \widehat{\mathbf{U}}}{\partial t}+\frac{\partial \widehat{\mathbf{E}}}{\partial \xi} = \mathbf{0}
    \label{eq:NS_cons_iso_1D}
\end{align}
with $\widehat{\mathbf{E}}=\widehat{\mathbf{E}}_{c}+\widehat{\mathbf{E}}_{d}$. The SD principle assumes that vector $\widehat{\mathbf{U}}$ varies as a polynomial of degree $p$ inside $\mathcal{L}$. Building a polynomial of degree $p$ for $\widehat{\mathbf{U}}$ requires $p+1$ points inside $\mathcal{L}$, called solution points (SP), where $\widehat{\mathbf{U}}$ has to be known. The location of SP is given by the Gauss-Chebyshev points of the first kind defined in~\cite{sun2007high} noted $\boldsymbol{\xi}_{SP}$. Consequently, $\boldsymbol{\xi}_{SP}$ is a vector of $p+1$ components which are the coordinates in segment $[0,1]$ of each SP inside $\mathcal{L}$. Because of Eq.~(\ref{eq:NS_cons_iso_1D}), the flux divergence should be a polynomial of degree $p$ too so that $\widehat{\mathbf{E}}$ has to be a polynomial of degree $p+1$. Then, $p+2$ points are required inside $\mathcal{L}$ to build a polynomial of degree $p+1$ for $\widehat{\mathbf{E}}$. These points are called flux points (FP) and are taken as the Gauss-Legendre points for the $p$ interior FP and the two remaining FP are placed at the boundaries of $\mathcal{L}$ namely $\xi=0$ and $\xi=1$~\cite{sun2007high}. This distribution is noted $\boldsymbol{\xi}_{FP}$ and is then a vector of $p+2$ components which are the coordinates in segment $[0,1]$ of each FP inside $\mathcal{L}$. 

\noindent At a given time, the solution vector $\widehat{\mathbf{U}}^{e}_{SP}=\left[\widehat{\mathbf{U}}^{e}_{i,SP}\right]_{1\leq i\leq N_{SP}}$ contains the conservative variables in the isoparametric domain stored at each solution point $i$ inside $\mathcal{L}$. This vector is used to construct a $p$-degree polynomial representation of a continuous solution $\mathbf{\overline{U}}$ across $\mathcal{L}$ using a Lagrange polynomial basis:
\begin{equation}
    \mathbf{\overline{U}}^{e}\left(\xi\right)= \sum\limits_{i=1}^{p+1}{\left[\widehat{\mathbf{U}}^{e}_{i,SP}L_{i,SP}\left(\xi\right)\right]} \hspace{0.25 cm}\text{for}\hspace{0.25 cm} \xi\in [0,1]
    \label{eq:continuous_U_1D}
\end{equation}
where $L_{i,SP}\left(\xi\right)$ is the Lagrange polynomial of degree $p$ based on SP of index $i$. Then, using the polynomial defined by Eq.~(\ref{eq:continuous_U_1D}) conservative variables are interpolated at internal FP ($2\leq i\leq p+1$) and extrapolated at interface FP ($i=1$ and $i=p+2$). At internal FP, flux vector is evaluated using successively  Eqs.~(\ref{eq:def_fluxes_src_terms}) and~(\ref{eq:def_E_F_G_hat}) to get $\left[\widehat{\mathbf{E}}^{e}_{i,FP}\right]_{2\leq i\leq p+1}$. At interface FP, a Riemann solver is used for the convective fluxes to obtain unique values for $\widehat{\mathbf{E}}_{c}$ noted $\widehat{\mathbf{E}}_{c}^{I}$. The Riemann solver employed during all this work is the HLLC solver~\cite{batten1997choice}. For the diffusive fluxes a centred scheme as in~\cite{sun2007high} is used to also have unique values for $\widehat{\mathbf{E}}_{d}$ noted $\widehat{\mathbf{E}}_{d}^{I}$. It means that the solution and the gradient used to compute viscous fluxes at the interface FP are computed using the arithmetic averages of the left and right solution and gradient at this interface FP. After this step, $\widehat{\mathbf{E}}^{e}_{1,FP}$ and $\widehat{\mathbf{E}}^{e}_{p+2,FP}$ will be known. From here, a $\left(p+1\right)$-degree polynomial representation of a continuous flux $\mathbf{\overline{E}}^{e}\left(\xi\right)$ across $\mathcal{L}$ can be constructed from flux values at FP:
\begin{equation}
    \mathbf{\overline{E}}^{e}\left(\xi\right) = \sum\limits_{i=1}^{p+2}{\left[\widehat{\mathbf{E}}^{e}_{i}L_{i,FP}\left(\xi\right)\right]}\hspace{0.25 cm}\text{for}\hspace{0.25 cm} \xi\in [0,1]
    \label{eq:continuous_E_1D}
\end{equation}
where $L_{i,FP}\left(\xi\right)$ is the Lagrange polynomial of degree $p+1$ based on FP of index $i$. The flux polynomial defined by Eq.~(\ref{eq:continuous_E_1D}) is differentiated along the $\xi$ direction and evaluated at solution points to obtain:
\begin{equation}
    \left(\frac{\partial \widehat{\mathbf{E}}}{\partial \xi}\right)^{e}_{SP} =  \sum\limits_{i=1}^{p+2}{\left[\widehat{\mathbf{E}}^{e}_{i}\frac{\partial L_{i,FP}}{\partial \xi}\left(\boldsymbol{\xi}_{SP}\right)\right]}
    \label{eq:flux_deri_1D}
\end{equation}
Finally, Eq.~(\ref{eq:NS_cons_iso_1D}) can be marched in time using any explicit temporal scheme at all SP inside all mesh elements:
\begin{equation}
   \frac{d \widehat{\mathbf{U}}^{e}_{SP}}{dt} = - \left(\frac{\partial \widehat{\mathbf{E}}}{\partial \xi}\right)^{e}_{SP}
    \label{eq:semi_disc_1D}
\end{equation}
In this work, the three stages and third-order in time total variation diminishing (TVD) Runge-Kutta (RK) scheme of Gottlieb and Shu is considered \cite{gottlieb1998total}. 

\subsection{Use of primitive variables at FP}
\label{sub:use_prim_var_FP}
\noindent As described in paragraph~\ref{sub:GenePrin1DSD}, the usual discretization process in SD considers an extrapolation of conservative variables from SP to FP. This approach will be refered as CONS. Nevertheless, another approach can be considered which is to compute $T$, $\mathbf{u}$, $P$ and $\mathbf{Y}$ at SP from conservative variables at SP and then extrapolate $T$, $\mathbf{u}$, $P$ and $\mathbf{Y}$ at FP. This approach can be refered as TUPY. Both approaches can be summed up as:
\begin{itemize}
    \item[$\bullet$] \textbf{CONS}: $\widehat{\mathbf{U}}^{e}_{SP}\rightarrow \widehat{\mathbf{U}}^{e}_{FP}$. $\widehat{\mathbf{U}}^{e}_{FP}$ is used at FP to compute any other needed variables for flux computations.
    \item[$\bullet$] \textbf{TUPY}: $\widehat{\mathbf{U}}^{e}_{SP}\rightarrow \widehat{\mathbf{Q}}^{e}_{SP}\rightarrow \widehat{\mathbf{Q}}^{e}_{FP}$. $\widehat{\mathbf{Q}}^{e}_{FP}$ is used at FP to compute any other needed variables for flux computations.
\end{itemize}
where $\mathbf{U}_{SP}=\left(\rho, \rho u, \rho v, \rho w, \rho E, \rho Y_{1},\hdots,\rho Y_{N_{s}}\right)^{\mathrm{T}}_{SP}$ and $\mathbf{Q}_{SP}=\left(T, u, v, w, P, Y_{1},\hdots,Y_{N_{s}}\right)^{\mathrm{T}}_{SP}$.
Actually, it was found that approach CONS is unstable, whereas approach TUPY is not, in the multispecies case while using the SD method. Before showing a pathological case, the way to compute temperature and pressure at any set of points from conservative variables at these points using JANAF enthalpy tables for a multispecies gas is explained in Algorithm~\ref{alg:Alg_CONS2T_and_P_multi}:

\begin{algorithm}[!h]
\caption{Compute temperature and pressure from conservative variables using JANAF enthalpy tables.}
\label{alg:Alg_CONS2T_and_P_multi}
\begin{algorithmic}[1]
\State Conservative variables in physical space $\mathbf{U}$ are known.
\State Compute sensible energy: $\rho e_{s} = \rho E - \frac{\left(\left(\rho u\right)^{2}+\left(\rho v\right)^{2}+\left(\rho w\right)^{2}\right)}{2\rho}$
\For{each interval $I_{n}=[T_{1}^{n},T_{2}^{n}]$, with $T_{2}^{n}-T_{1}^{n}=100\ \mathrm{K}$, of JANAF tables}
    \State Compute: $\rho e_{s}\left(T_{1}^{n}\right) = \sum\limits_{k=1}^{N_{s}}{\rho Y_{k}e_{sk}\left(T_{1}^{n}\right)}\hspace{0.25 cm}\text{and}\hspace{0.25 cm} \rho e_{s}\left(T_{2}^{n}\right) = \sum\limits_{k=1}^{N_{s}}{\rho Y_{k}e_{sk}\left(T_{2}^{n}\right)}$
    \If{$\rho e_{s}\left(T_{1}^{n}\right)\leq \rho e_{s}\leq \rho e_{s}\left(T_{2}^{n}\right)$}
        \State $T = \left(n + \frac{\rho e_{s} - \rho e_{s}\left(T_{1}^{n}\right)}{\rho e_{s}\left(T_{2}^{n}\right) - \rho e_{s}\left(T_{1}^{n}\right)}\right)\times 100$
    \Else
        \State Go to $I_{n+1}$ and do the same test again until $\rho e_{s}\left(T_{1}^{n}\right)\leq \rho e_{s}\leq \rho e_{s}\left(T_{2}^{n}\right)$ is verified and deduced $T$ from it.
    \EndIf
\EndFor
\State At this point, $T$ is known.
\State Compute $r \equiv C_{p}\left(T\right)-C_{v}\left(T\right) = \sum\limits_{k=1}^{N_{s}}{\left(\rho Y_{k}\right)\left[C_{pk}\left(T\right)-C_{vk}\left(T\right)\right]/\rho}=\frac{\overline{R}}{W}$ and then $P$ using Eq.~(\ref{eq:perfect_gas_law}).  
\end{algorithmic}
\end{algorithm}
\noindent Sensible energies and heat capacities at both constant pressure and volume are computed from JANAF sensible enthalpy tables, for each species $k$, as:
\begin{align}
    e_{sk}\left(T_{1}^{n}\right) &= h_{sk}\left(T_{1}^{n}\right) - \frac{\overline{R}T_{1}^{n}}{W_{k}}\hspace{0.10 cm}\text{,}\hspace{0.10 cm} C_{pk}\left(T_{1}^{n}\right) =\frac{h_{sk}\left(T_{2}^{n}\right)-h_{sk}\left(T_{1}^{n}\right)}{T_{2}^{n}-T_{1}^{n}}\hspace{0.10 cm}\text{,}\hspace{0.10 cm} C_{vk}\left(T_{1}^{n}\right) =\frac{e_{sk}\left(T_{2}^{n}\right)-e_{sk}\left(T_{1}^{n}\right)}{T_{2}^{n}-T_{1}^{n}}
\end{align}
\noindent Thus, for the multispecies case, approaches CONS and TUPY are not equivalent. The first one uses Algorithm~\ref{alg:Alg_CONS2T_and_P_multi} at FP with $\mathbf{U}=\mathbf{U}_{FP}$ as input, whereas the second one uses Algorithm~\ref{alg:Alg_CONS2T_and_P_multi} at SP with $\mathbf{U}=\mathbf{U}_{SP}$ and then do the extrapolation of $\mathbf{Q}$ at FP. 

\begin{figure}[!h]
    \centering
    \includegraphics[width=\Lfig, height=\Hfig]{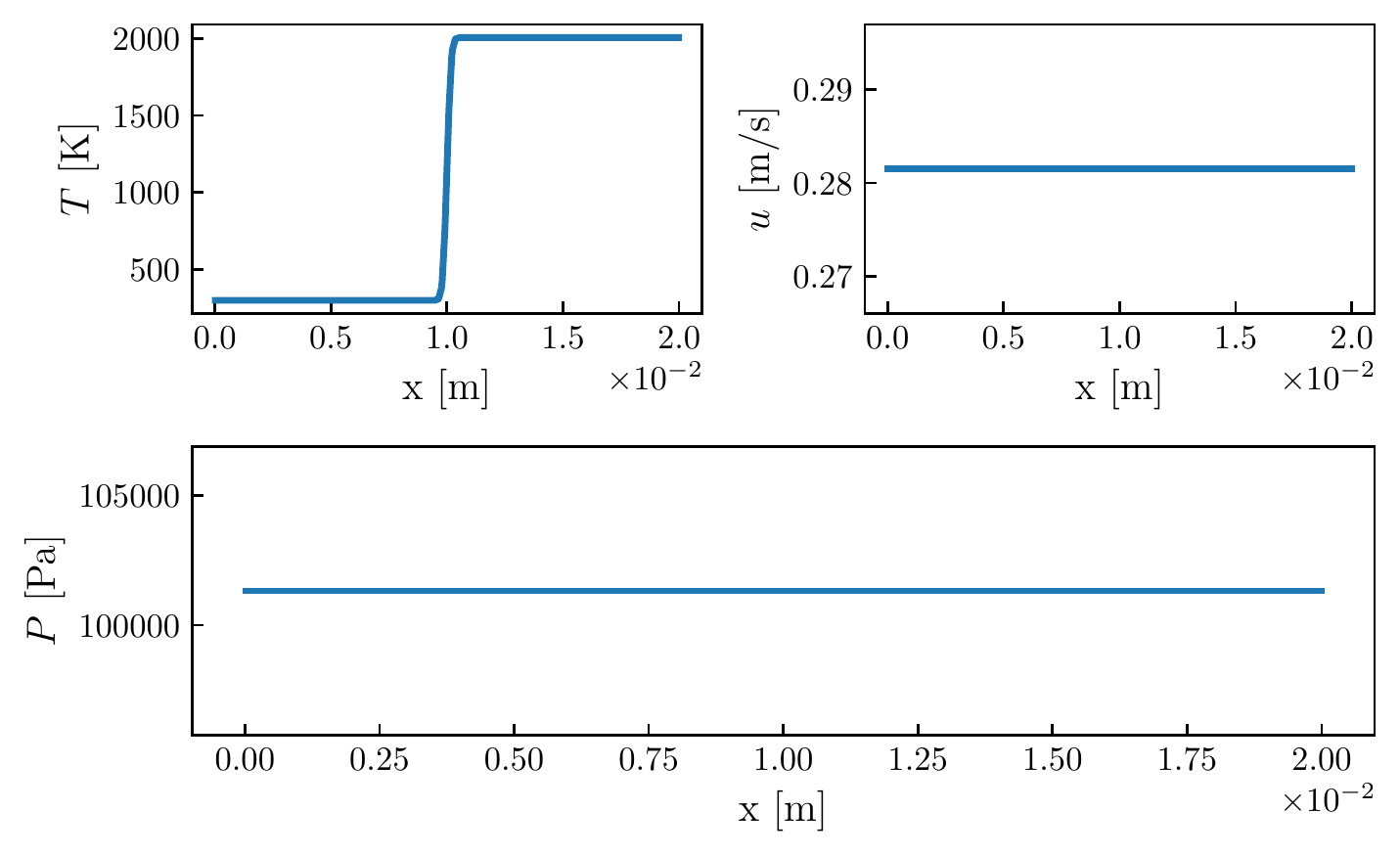}
    \caption{Initial values of $T$, $u$ and $P$ at SP for both CONS and TUPY approaches.}
    \label{fig:initial_TUP_SP}
\end{figure}

\noindent To highlight this non-equivalence, a one-dimensional domain with fresh gases on the left side at $T=300$ K and burnt gases on the right side at $T=2010$ K is considered. Initial pressure and velocity are set constant respectively at $P=101325$ Pa and $u=0.2815\ \mathrm{m.s^{-1}}$ along with global equivalence ratio set to 0.8. This is the initial situation presented in Table~\ref{tab:1D_BFER_phi08_charac} and illustrated on Figure~\ref{fig:initial_TUP_SP} where values of $T$, $u$ and $P$ at SP are shown. Characteristic boundary conditions for a subsonic inlet imposing $T_{in}=300$ K, $u_{in}=0.2815\ \mathrm{m.s^{-1}}$ and mass fractions to have $\phi=0.8$ and for a subsonic outlet imposing $P_{out}=101325$ Pa are used. The way to impose these boundary conditions within the SD context is described in Section~\ref{sec:charac_bc_SD}.  

\begin{figure}[!h]
    \centering
    \includegraphics[width=\Lfig, height=\Hfig]{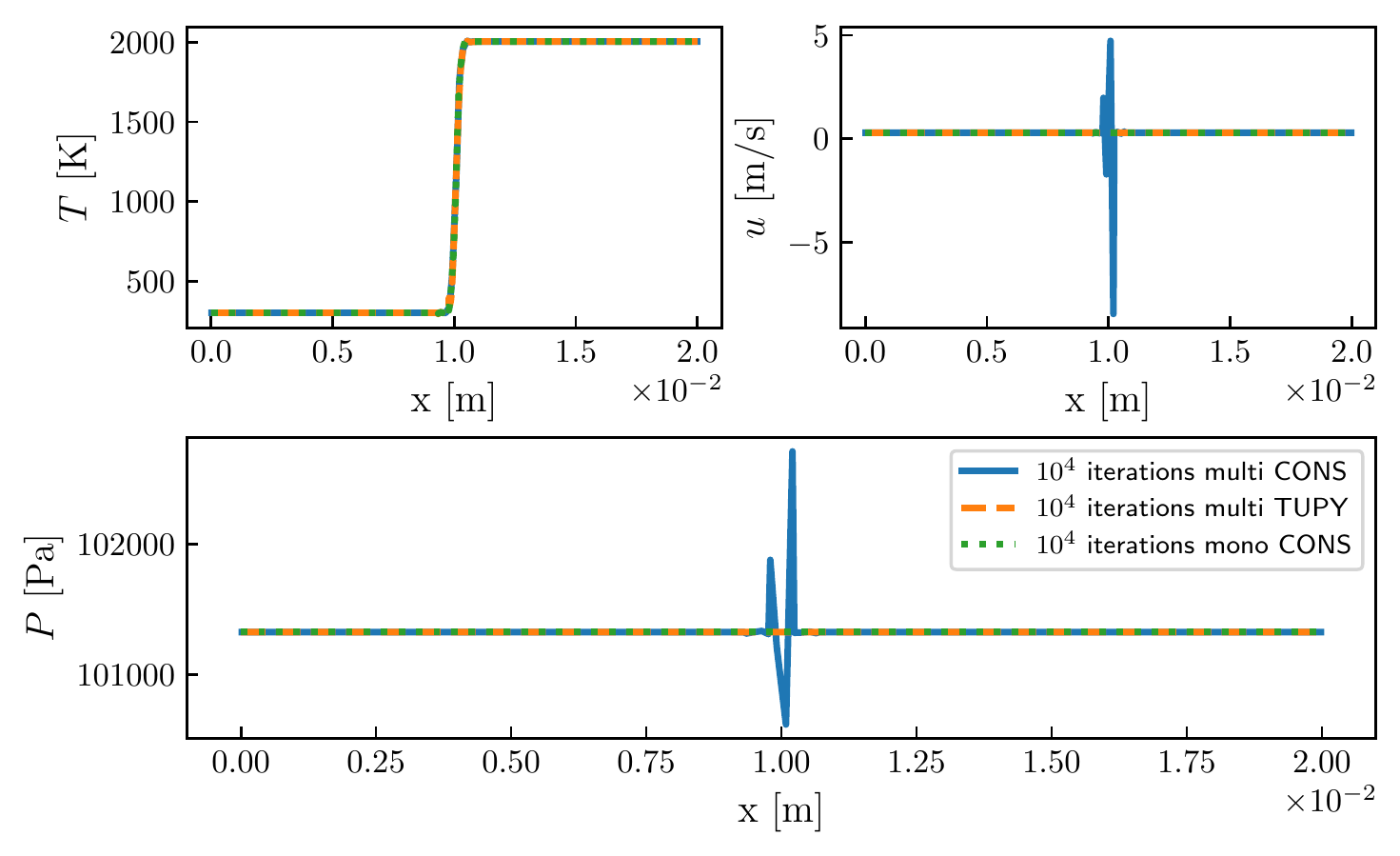}
    \caption{Values of $T$, $u$ and $P$ at SP after $10^{4}$ iterations for multispecies cases with either CONS or TUPY approaches and monospecies case with CONS approach.}
    \label{fig:TUP_SP_10000_multi_mono}
\end{figure}

\noindent Then, the simulation is runned without diffusion fluxes (to only see convective effects) and also without combustion source terms activated. This is the situation of a contact discontinuity with only temperature and species mass fractions (and so the density) that are changing in the transition zone between fresh and burnt gases. In that case, pressure and velocity must remain constant during the computation. Unfortunately, this is not the case for CONS approach where wiggles appear on the pressure (and consequently on velocity too) in the transition zone between fresh and burnt gases as depicted in Figure~\ref{fig:TUP_SP_10000_multi_mono}. On the contrary, it can be seen that these wiggles are not present for the TUPY approach and also when CONS approach is used for a monospecies gas with constant thermodynamic properties in the same situation (so with a temperature and density gradient).

\begin{figure}[!h]
    \centering
    \includegraphics[width=\Lfig, height=\Hfig]{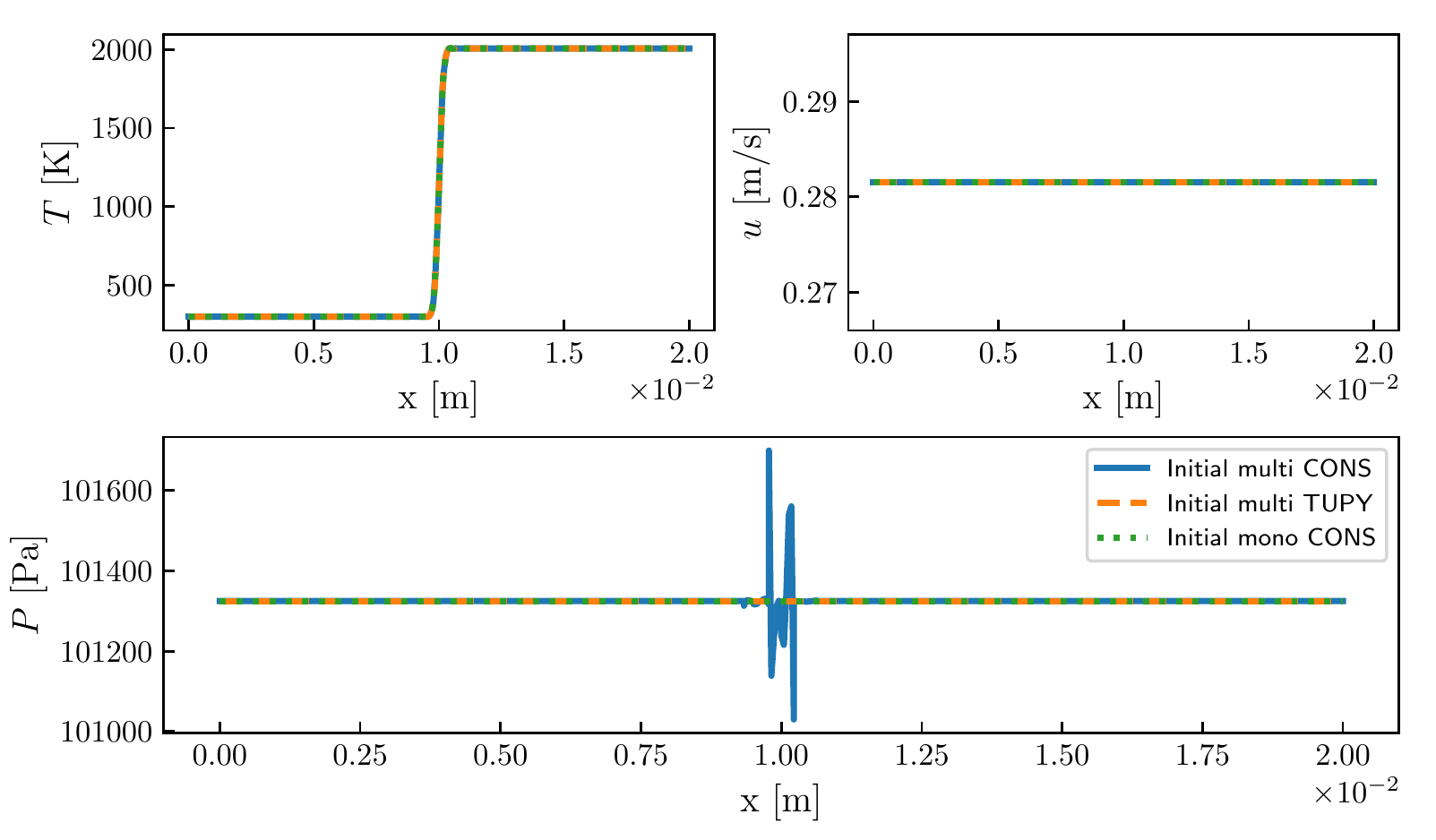}
    \caption{Initial values of $T$, $u$ and $P$ at OP for multispecies cases with either CONS or TUPY approaches and monospecies case with CONS approach.}
    \label{fig:initial_TUP_OP_multi_mono}
\end{figure}

\noindent To better understand what is happening, values of $T$, $u$ and $P$ can be represented at the initialization (when no time iterations were done) on ten points equally spaced per mesh element, called output points (OP), for both the monospecies case and the multispecies case. This is done to see values of $T$, $u$ and $P$ on other points than SP from which their continous polynomials are built. It is presented in Figure~\ref{fig:initial_TUP_OP_multi_mono} where it can be seen that pressure has large wiggles, directly at initialization, in the transition zone between fresh and burnt gases when CONS approach is used whereas there are no wiggles for the two other cases. Therefore, for the multispecies case, it seems that building polynomials of conservative variables, evaluate them on another points than SP and compute $P$ at theses points create pressure oscillations. On the other hand, building polynomials of $P$ and $T$ at SP and evaluate them on another points than SP does not create pressure oscillations. That is why, for a flow composed of a multispecies gas simulated using the SD method, the TUPY method is prefered which is the case in this work. The reason why is complex to write mathematically but it probably comes from the fact that for a multispecies gas, pressure and total energy are not linked following a linear relation such as the one for a calorically perfect gas:
\begin{align}
P = \left(\gamma-1\right)\left(\rho E- \frac{\left(\left(\rho u\right)^{2}+\left(\rho v\right)^{2}+\left(\rho w\right)^{2}\right)}{2\rho}\right)
    \label{eq:compute_press_mono}
\end{align}
where the heat capacity ratio $\gamma$ is constant and not temperature and species dependent as for a multispecies gas. Indeed, for a multispecies gas, pressure and total energy are linked according to Eq.~(\ref{eq:def_rhoE_multi_thermally}) which is completely non-linear since sensible energy $e_{s}$ is temperature and species dependent. Then, $\gamma$ is non-constant and without any treatments, a pressure jump can appear through any contact discontinuity (such as the one initialized in Figure~\ref{fig:initial_TUP_SP}) and this jump will create spurious oscillations of all physical quantities according to Abgrall et al.~\cite{abgrall2001computations}. The same authors have proposed a special treatment named as the \textit{Double Flux method} which has been successfully applied in reactive DG computations~\cite{billet2014runge,lv2014discontinuous,lv2017high}. Adapting this treatment to the SD formalism is out of the scope of this paper but will be the purpose of future work.

\noindent It should be mentioned that another approach extrapolating $\rho$ instead of $T$ at FP, called $\rho$UPY was also tested. It was also much more stable than CONS approach and results were similar to the ones of TUPY.

%% file: 5_NSCBCBoundaryConditions.tex
\section{Characteristic boundary conditions for SD\label{sec:charac_bc_SD}}
\noindent Having accurate boundary conditions is essential to simulate combustion applications. Characteristic boundary conditions are widely employed in this context but their use with the SD method is very recent. The objective of this section is to explain how characteristic boundary conditions can be implemented and used along with the SD method applied to the reacting compressible NSE for a multispecies gas. As it was done by Baum et al.~\cite{baum1995accurate} who extended the work of Poinsot and Lele~\cite{poinsot1992boundary} in multicomponent reactive flows for a cartesian coordinate system, this is an extension of the work of Fievet et al.~\cite{fievet2020strong} to multicomponent reactive flows solved in a generalized coordinate system.

\subsection{Useful formula}
\label{sub:useful_formula_nscbc}
\begin{figure}[!h]
    \centering
    \includegraphics[width=9cm]{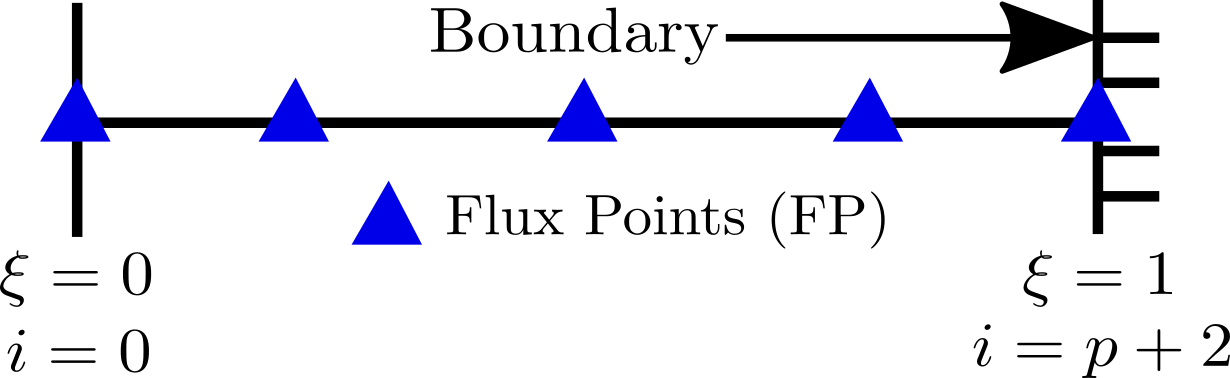}
    \caption{Illustration of the computation of a function $f$ at a boundary FP to impose $\left(\partial f/\partial \xi\right)$ at this boundary FP. The polynomial degree is set to $p=3$ with five FP along direction $\xi$.}
    \label{fig:boundary_general_case}
\end{figure}

\noindent It will be shown in the next parts that for particular boundary conditions, derivatives of some scalars or vectors have to be imposed. However, in this work it was chosen to not directly impose the derivatives but rather to compute the scalar or vector value that will satisfy the condition on its derivative. Let's denote by $f$ any scalar (pressure, temperature or mass fraction) or vector (flux vector) function. Without a loss of generality, if the boundary is at $\xi=cste$ in the isoparametric domain, $f$ can be differentiated with respect to $\xi$ (normal direction of the boundary) using derivatives of Lagrange polynomials at FP:
\begin{equation}
    \frac{\partial f}{\partial \xi}\left(\xi\right) = \sum\limits_{i=1}^{p+2}{f_{i,FP}\frac{\partial L_{i,FP}}{\partial \xi}\left(\xi\right)}\label{eq:dfdxi_lagrange}
\end{equation}
where $f_{i,FP}$ is the value of $f$ at FP $i$ in direction $\xi$. This situation is represented in Figure~\ref{fig:boundary_general_case} for a boundary located at $\xi=1$ when the polynomial degree is $p=3$ in the element close to the boundary (five FP along $\xi$ direction).
In that case, $\left(\partial f/\partial \xi\right)\left(1\right)$ has to be imposed to a known value according to the type of boundary condition considered. This implies that $f$ at $\xi=1$, which is $f_{p+2,FP}$, is computed by inverting Eq.~(\ref{eq:dfdxi_lagrange}) to get:
\begin{equation}
    f_{p+2,FP} = \frac{\frac{\partial f}{\partial \xi}\left(1\right) - \sum\limits_{i=1}^{p+1}{f_{i,FP}\frac{\partial L_{i,FP}}{\partial \xi}\left(1\right)}}{\frac{\partial L_{p+2,FP}}{\partial \xi}\left(1\right)} 
    \label{eq:get_fpPLUS2_hat_FP}
\end{equation}
\noindent It should be mentioned that the value of $f$ at the opposite interface FP, here $f_{1,FP}$, is a continuous value of $f$ at $\xi=0$ interface separating the element that contains the boundary and its closest neighbor along $\xi$ direction. Then, for scalar values of $f$, $f_{1,FP}$ is taken as the average value of $f$ at $\xi=0$ interface and for flux vectors it is the interface flux after Riemann solver and diffusion scheme have been applied. The same principle can be used for a boundary located at $\xi=0$, where $\left(\partial f/\partial \xi\right)\left(0\right)$ has to be imposed, so that $f$ at $\xi=0$ must be given by Eq.~(\ref{eq:get_f1_hat_FP}):
\begin{equation}
    f_{1,FP} = \frac{\frac{\partial f}{\partial \xi}\left(0\right) - \sum\limits_{i=2}^{p+2}{f_{i,FP}\frac{\partial L_{i,FP}}{\partial \xi}\left(0\right)}}{\frac{\partial L_{1,FP}}{\partial \xi}\left(0
    \right)} 
    \label{eq:get_f1_hat_FP}
\end{equation}
More generally, this reasoning is exactly the same for boundaries along the other directions in the isoparametric domain.

\subsection{Mathematical framework}

\subsubsection{Specificity of the SD method}
\label{subsub:specifity_SD}
\noindent In a recent paper, Fievet et al.~\cite{fievet2020strong} have shown how to implement Navier-Stokes characteristic boundary conditions (NSCBC) with a SD discretization. Let's consider without loss of generality a boundary located at a $\xi$-normal face in $\xi=1$ as in Figure~\ref{fig:boundary_general_case}. The idea of the NSCBC treatment is to impose a value of $\left(\partial \widehat{\mathbf{E}}/\partial \xi\right)$ at the boundary FP in $\xi=1$
according to the waves crossing this boundary. However as shown in Eq.~(\ref{eq:semi_disc_1D}), the flux derivatives are computed at SP from the flux polynomial built at the FP. Thus, the NSCBC approach will actually provides the expected value of $\left(\partial \widehat{\mathbf{E}}/\partial \xi\right)$ at the FP in $\xi=1$ that will be used to deduce a flux value at this FP. Let's denote by $\left(\partial \widehat{\mathbf{E}}/\partial \xi\right)^{*}$ and $\widehat{\mathbf{E}}^{*}$ respectively the corrected flux derivative and flux values obtained after the NSCBC treatment applied in $\xi=1$. The flux at the boundary FP to be imposed is given by Eq.~(\ref{eq:get_fpPLUS2_hat_FP}) with $f=\widehat{\mathbf{E}}$:
\begin{align}
    \widehat{\mathbf{E}}_{p+2,FP}^{*}=\frac{\frac{\partial \widehat{\mathbf{E}}}{\partial \xi}^{*}\left(1\right)-\sum\limits_{i=1}^{p+1}{\widehat{\mathbf{E}}_{i,FP}\frac{\partial L_{i,FP}}{\partial \xi}\left(1\right)}}{\frac{\partial L_{p+2,FP}}{\partial \xi}\left(1\right)} \label{eq:Ehat_xi_1}
\end{align}
After this treatment, the value of $\left(\partial \widehat{\mathbf{E}}/\partial \xi\right)$ at SP will be corrected into~\cite{fievet2020strong}:
\begin{align}
    \frac{\partial \widehat{\mathbf{E}}}{\partial \xi}^{*}\left(SP\right) = \left[\frac{\partial \widehat{\mathbf{E}}}{\partial \xi}^{*}\left(1\right)-\frac{\partial \widehat{\mathbf{E}}}{\partial \xi}\left(1\right)\right]\frac{\partial_{\xi} L_{p+2,FP}\left(SP\right)}{\partial_{\xi} L_{p+2,FP}\left(1\right)} + \frac{\partial \widehat{\mathbf{E}}}{\partial \xi}\left(SP\right)
\end{align}
Consequently, $\left(\partial \widehat{\mathbf{E}}/\partial \xi\right)^{*}\left(SP\right)$ is the modified value of $\left(\partial \widehat{\mathbf{E}}/\partial \xi\right)$ that will be used in Eq.~(\ref{eq:semi_disc_1D}) to march $\widehat{\mathbf{U}}$ in time. 

\subsubsection{Wave equation}
\label{subsub:wave_eqn_nscbc}
\noindent The determination of the expected flux derivative at the FP at $\xi=1$ on the NSCBC starts with the diagonalization of $\left(\partial \widehat{\mathbf{E}}/\partial \xi\right)$ in Eq.~(\ref{eq:NS_cons_iso_3rd}) which gives the following wave equation \cite{fievet2020strong}:
\begin{equation}
    |J|\frac{\partial \mathbf{W}}{\partial t} + \boldsymbol{\mathcal{N}} = -\boldsymbol{\mathcal{S}}
    \label{eq:characteristic_equation_fievet}
\end{equation}
where $\mathbf{W}$, $\boldsymbol{\mathcal{N}}$ and $\boldsymbol{\mathcal{S}}$ are respectively the vector of characteristic variables, the normal and tangential strengths of the characteristics given by:
\begin{align}
    \partial \mathbf{W} &= P_{\mathbf{U}}\partial \mathbf{U} \label{eq:Def_partialW}\\
    \boldsymbol{\mathcal{N}} &\equiv
    \left(
    \begin{array}{c}
        \mathcal{N}_{1}   \\
        \mathcal{N}_{2} \\
        \mathcal{N}_{3} \\
        \mathcal{N}_{+} \\
        \mathcal{N}_{-} \\
        \mathcal{N}_{5+1} \\
        \vdots \\
        \mathcal{N}_{5+N_{s}}
    \end{array}
    \right)
     = P_{\mathbf{U}}\left(\frac{\partial \widehat{\mathbf{E}}}{\partial \xi}-\mathbf{\mathcal{A}}_{c}\left(\xi\right)-\mathbf{\mathcal{A}}_{d}\left(\xi\right)\right) \label{eq:Def_N_nscbc}\\
    \boldsymbol{\mathcal{S}} &\equiv
    \left(
    \begin{array}{c}
        \mathcal{S}_{1}   \\
        \mathcal{S}_{2} \\
        \mathcal{S}_{3} \\
        \mathcal{S}_{+} \\
        \mathcal{S}_{-} \\
        \mathcal{S}_{5+1} \\
        \vdots \\
        \mathcal{S}_{5+N_{s}}
    \end{array}
    \right)
    =P_{\mathbf{U}}\left(\frac{\partial \widehat{\mathbf{F}}}{\partial \eta} + \frac{\partial \widehat{\mathbf{G}}}{\partial \zeta} + \mathbf{\mathcal{A}}_{c}\left(\xi\right) + \mathbf{\mathcal{A}}_{d}\left(\xi\right)\right)  \label{eq:Def_S_nscbc}
\end{align}
with $\mathbf{\mathcal{A}}_{c}\left(\xi\right)$ and $\mathbf{\mathcal{A}}_{d}\left(\xi\right)$ terms accounting for the mesh non-orthogonality \cite{fievet2020strong}:
\begin{align}
    \mathbf{\mathcal{A}}_{c/d}\left(\xi\right) = \mathbf{E}_{c/d}\frac{\partial}{\partial \xi}\left(\xi_{x}|J|\right)+\mathbf{F}_{c/d}\frac{\partial}{\partial \xi}\left(\xi_{y}|J|\right)+\mathbf{G}_{c/d}\frac{\partial}{\partial \xi}\left(\xi_{z}|J|\right)
\end{align}
In Eqs.~(\ref{eq:Def_partialW}-\ref{eq:Def_S_nscbc}),  $P_{\mathbf{U}}$ is the transformation matrix from conservative to characteristic variables usually expressed following:
\begin{equation}
   P_{\mathbf{U}} = P_{\mathbf{Q}}.\frac{\partial \mathbf{Q}}{\partial \mathbf{U}}
   \label{eq:link_PU_PQ}
\end{equation}
\begin{equation}
    P_{\mathbf{Q}} \equiv \frac{\partial \mathbf{W}}{\partial \mathbf{Q}} = \left[
    \begin{array}{cccccc}
       n_{x} & 0 & n_{z} & -n_{y} & -n_{x}/c^{2} & O_{1,N_{s}} \\
        n_{y} & -n_{z} & 0 & n_{x} &-n_{y}/c^{2} & O_{1,N_{s}} \\
        n_{z} & n_{y} & -n_{x} & 0 & -n_{z}/c^{2} & O_{1,N_{s}} \\
        0 & n_{x}/\sqrt{2} & n_{y}/\sqrt{2} & n_{z}/\sqrt{2} & 1/\left(\sqrt{2}\rho c\right) & O_{1,N_{s}} \\
        0 & -n_{x}/\sqrt{2} & -n_{y}/\sqrt{2} & -n_{z}/\sqrt{2} & 1/\left(\sqrt{2}\rho c\right) & O_{1,N_{s}} \\
        O_{N_{s},1} & O_{N_{s},1} & O_{N_{s},1} & O_{N_{s},1} & O_{N_{s},1} & I_{N_{s},N_{s}}
    \end{array}
    \right]
    \label{eq:def_PQ}
\end{equation}
\begin{equation}
    \frac{\partial \mathbf{Q}}{\partial \mathbf{U}}= \left[
    \begin{array}{cccccccc}
        1 & 0 & 0 & 0 & 0 & O_{1,N_{s}} & &\\
        \frac{-u}{\rho} & \frac{1}{\rho} & 0 & 0 & 0 & O_{1,N_{s}} & &\\
        \frac{-v}{\rho} & 0 & \frac{1}{\rho} & 0 & 0 & O_{1,N_{s}} & &\\
        \frac{-w}{\rho} & 0 & 0 & \frac{1}{\rho} & 0 & O_{1,N_{s}} & &\\
        \frac{\left(\gamma-1\right)||\mathbf{u}||^{2}}{2} & \left(1-\gamma\right)u & \left(1-\gamma\right)v & \left(1-\gamma\right)w &\gamma -1& \frac{\partial P}{\partial \rho Y_{1}} & \hdots & \frac{\partial P}{\partial \rho Y_{N_{s}}}\\
        \frac{-Y_{1}}{\rho} & O_{N_{s},1} & O_{N_{s},1} & O_{N_{s},1} & O_{N_{s},1} &  & & \\
        \vdots &  &  &  &  & & \frac{1}{\rho}I_{N_{s},N_{s}} & \\
        \frac{-Y_{N_{s}}}{\rho} &  &  &  &  & & &
    \end{array}
    \right]
    \label{eq:dQdU_matrix}
\end{equation}
where $O_{m,n}$ is the zero matrix of dimension $m\times n$, $I_{N_{s},N_{s}}$ is the identity matrix of size $N_{s}$, $\mathbf{n}=\left(n_{x},n_{y},n_{z}\right)^{\mathrm{T}}$ is the face unit normal vector (the normal at a given FP for instance) taken as $\nabla \xi/||\nabla \xi||$ and $c$ is the local speed of sound. Additionally, in Eq.~(\ref{eq:dQdU_matrix}) $\left(\partial P/\partial \rho Y_{k}\right)$ is given by:
\begin{align}
\frac{\partial P}{\partial \rho Y_{k}}=\left(1-\gamma\right)\left(h_{sk}-TC_{p}\frac{W}{W_{k}}\right)\hspace{0.20 cm}\text{for}\hspace{0.20 cm} k=1,N_{s}
\end{align}
Vector $\partial \mathbf{W}$ is composed of $3+N_{s}$ entropy waves namely $\partial W_{1}$, $\partial W_{2}$, $\partial W_{3}$, $\partial W_{5+1},\hdots ,\partial W_{5+N_{s}}$, propagating at $u_{n}=\mathbf{u}.\mathbf{n}=un_{x}+vn_{y}+wn_{z}$ and of two acoustic waves $\partial W_{+}$ and $\partial W_{-}$ propagating respectively at $u_{n}+c$ and $u_{n}-c$. In Eqs.~(\ref{eq:link_PU_PQ}-\ref{eq:dQdU_matrix}), $\mathbf{Q}=\left(\rho,u,v,w,P,Y_{1},\hdots,Y_{N_{s}}\right)^{\mathrm{T}}$ has not to be confused with the set of primitive variables used here at FP as explained in paragraph \ref{sub:use_prim_var_FP} where density is replaced by temperature as first variable. Moreover, the transformation matrix $\left(\partial \mathbf{Q}/\partial \mathbf{U}\right)$ was extended to a multispecies thermally perfect gas compared to the work of Fievet et al.~\cite{fievet2020strong} considering a monospecies calorically perfect gas. This computation is given in~\ref{appendix:get_dQdU}. It should be mentioned that the heat release rate and species source terms are not taken into account at the NSCBC since the cases considered here do not show flames close to a NSCBC. This will be the objective of future work. 

\subsubsection{General algorithm for a NSCBC treatment using SD}
\noindent Following paragraphs \ref{subsub:specifity_SD} and \ref{subsub:wave_eqn_nscbc}, prescribing characteristic boundary conditions using NSE written in a generalized coordinates system can be summed up in Algorithm \ref{alg:Fievet_chara_bnd}.
\begin{algorithm}[!h]
\caption{Fievet et al. algorithm to do characteristic boundary conditions in generalized coordinates}
\label{alg:Fievet_chara_bnd}
\begin{algorithmic}[1]
\State Evaluate initial guesses for $\boldsymbol{\mathcal{N}}$ and $\boldsymbol{\mathcal{S}}$ at NSCBC FP in $\xi=1$ using Eqs.~(\ref{eq:Def_N_nscbc}) and (\ref{eq:Def_S_nscbc}).
\State Modify some values of $\boldsymbol{\mathcal{N}}$ to account for the NSCBC treatment. This gives $\boldsymbol{\mathcal{N}}^{*}$.
\State Compute new values of $\frac{\partial \widehat{\mathbf{E}}}{\partial \xi}\left(1\right)$ using Eq.~(\ref{eq:Def_N_nscbc}) with $\boldsymbol{\mathcal{N}}^{*}$. This gives $\frac{\partial \widehat{\mathbf{E}}^{*}}{\partial \xi}\left(1\right)$.
\State Compute the new flux value $\widehat{\mathbf{E}}_{p+2,FP}$ using Eq.~(\ref{eq:Ehat_xi_1}). This gives $\widehat{\mathbf{E}}_{p+2,FP}^{*}$.
\State Compute the new value of $\frac{\partial \widehat{\mathbf{E}}^{*}}{\partial \xi}\left(SP\right)$ which will be put back into Eq.~(\ref{eq:NS_cons_iso_3rd}) to march $\widehat{\mathbf{U}}$ in time.
\end{algorithmic}
\end{algorithm}
This methodology was successfully tested by Fievet et al.~\cite{fievet2020strong} on 1D and 2D Euler test cases such as one-dimensional acoustic wave or the convection of a 2D vortex. However, it was never employed with viscous and multispecies gas which is the purpose of this work.

\subsection{Types of NSCBC}
\label{sub:types_nscbc}
\begin{figure}[!h]
    \centering
    \includegraphics[width=\Lfig,height=\Hfig]{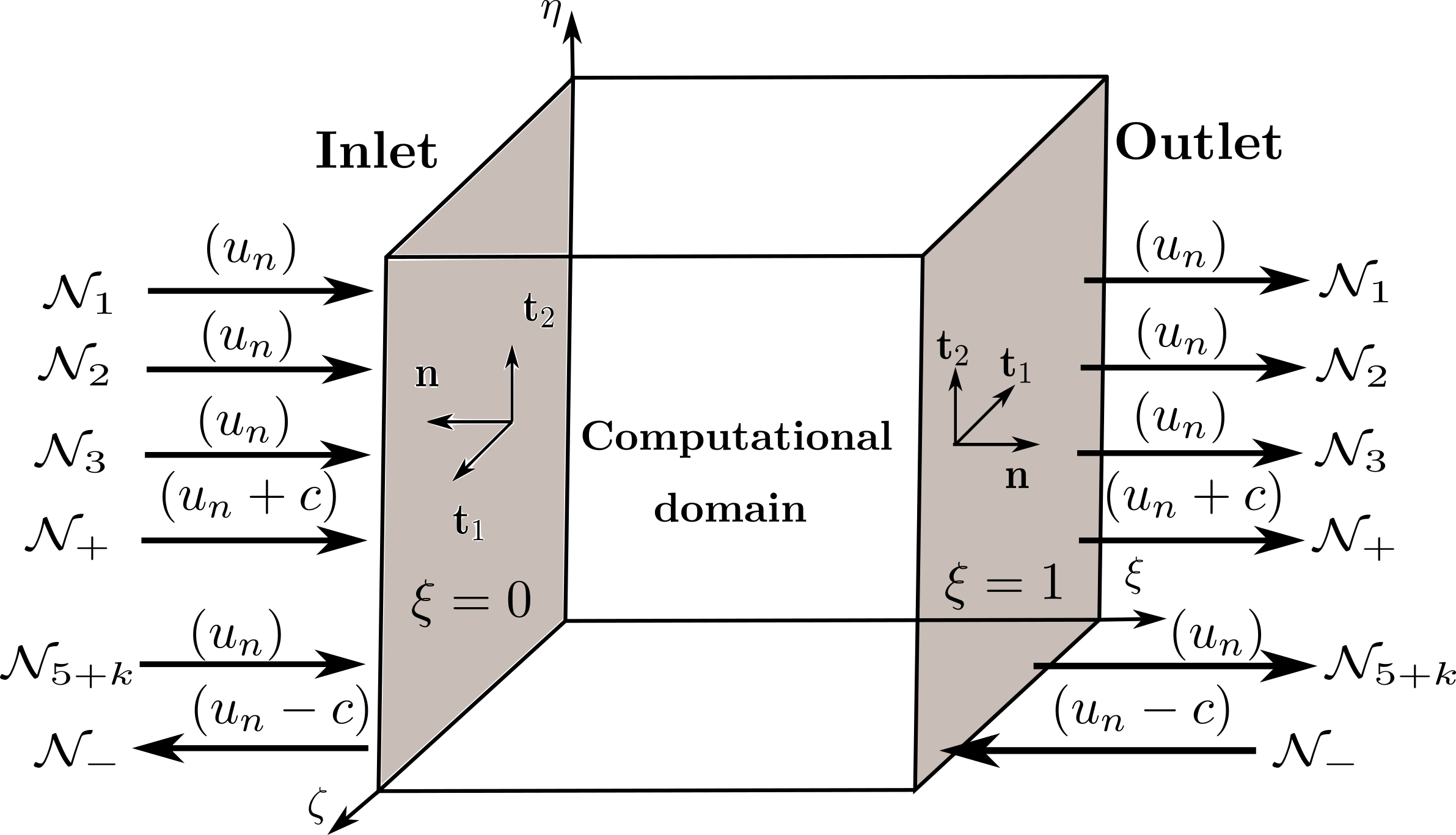}
    \caption{Waves crossing NSCBC inlet and outlet.}
    \label{fig:inlet_outlet_NSCBC}
\end{figure}

\noindent The objective of this section is to explain several types of NSCBC implemented in step 2 of Algorithm~\ref{alg:Fievet_chara_bnd}. A subsonic outflow and a subsonic inflow are considered along a $\xi$-normal boundary as shown in Figure~\ref{fig:inlet_outlet_NSCBC} with their associated waves. 


\subsubsection{Subsonic outflow imposing a constant pressure}
\noindent According to Figure~\ref{fig:inlet_outlet_NSCBC}, for a subsonic outflow, $4+N_{s}$ characteristic waves are leaving the domain (1 at speed $u_{n}+c$ and $3+N_{s}$ at speed $u_{n}$) while $1$ characteristic wave is entering the domain at speed $u_{n}-c$. Consequently, only one thermodynamic relation must be imposed to have a numerically well-behaved boundary condition~\cite{poinsot1992boundary}. Then, the entering wave is specified as~\cite{fievet2020strong}: 
\begin{equation}
    \mathbf{\mathcal{N}}^{*}_{-} = |J|K_{P}\left(P-P^{t}\right) + \alpha\mathbf{\mathcal{S}}_{-}^{\text{exact}} - \left(1-\alpha\right)\mathbf{\mathcal{S}}_{-}
    \label{eq:partially_NRO_transverse_term_Yoo}
\end{equation}
where $K_{P}\ \left[s^{-1}\right]$ is the pressure relaxation rate, $P^{t}$ and $P$ are respectively the target and current pressure at the boundary and $\alpha$ is a relaxation parameter usually taken as the averaged bulk Mach number over the whole boundary. In Eq.~(\ref{eq:partially_NRO_transverse_term_Yoo}), $\mathbf{\mathcal{S}}_{-}^{\text{exact}}$ is the exact value of $\mathbf{\mathcal{S}}_{-}$ that can be sometimes obtained using Eq.~(\ref{eq:Def_S_nscbc}) in some analytical test cases where mathematical expressions of $\left(\partial \widehat{\mathbf{F}}/\partial \eta\right)$ and $\left(\partial \widehat{\mathbf{G}}/\partial \zeta\right)$ are known~\cite{yoo2005characteristic,yoo2007characteristic}. It is always set to zero in this work. Note that with this formulation, the pressure is weakly imposed to limit undesired noise. At this point, only convective fluxes have been treated. According to~\cite{poinsot1992boundary}, for a subsonic outflow diffusive fluxes have to satisfy (in a case of a $\xi$-normal boundary):
\begin{align}
    \frac{\partial \left[\left(\boldsymbol{\tau}.\mathbf{n}\right).\mathbf{t}_{1}\right]}{\partial \xi} = 0\hspace{0.10 cm}\text{,}\hspace{0.10 cm} \frac{\partial \left[\left(\boldsymbol{\tau}.\mathbf{n}\right).\mathbf{t}_{2}\right]}{\partial \xi} = 0\hspace{0.10 cm}\text{,}\hspace{0.10 cm} \frac{\partial \mathbf{q}.\mathbf{n}}{\partial \xi} = 0\hspace{0.10 cm}\text{,}\hspace{0.10 cm}
    \frac{\partial \mathbf{M}_{k}.\mathbf{n}}{\partial \xi} = 0\hspace{0.20 cm}\text{for}\hspace{0.20 cm} k=1,N_{s} \label{eq:viscous_relations_nscbc_outlet}
\end{align}
where $\mathbf{t_{1}}$ and $\mathbf{t_{2}}$ are the two unit tangential vectors in the exit plane whose expressions are given in~\ref{appendix:tangential_vectors}. Derivatives in Eq.~(\ref{eq:viscous_relations_nscbc_outlet}) are imposed following the method described in paragraph \ref{sub:useful_formula_nscbc}. As they concern diffusive fluxes, $\mathbf{\widehat{E}}_{d}$, $\mathbf{\widehat{F}}_{d}$ and $\mathbf{\widehat{G}}_{d}$ are modified to take into account these conditions before the computation of $\widehat{\mathbf{E}}_{d}$ which is needed to get $\boldsymbol{\mathcal{N}}$ through Eq.~(\ref{eq:Def_N_nscbc}). 

\subsubsection{Subsonic inflow imposing velocities, temperature and species mass fractions}
\noindent According to Figure~\ref{fig:inlet_outlet_NSCBC}, for a subsonic inflow, $4+N_{s}$ characteristic waves are entering the domain while the remaining one exits it. Then, there are now $4+N_{s}$ thermodynamic relations to impose for a numerically well-behaved boundary condition~\cite{poinsot2005theoretical}. Let's say that $u$, $v$, $w$, $T$, $Y_{1},\hdots,Y_{N_{s}}$ are prescribed at the inflow boundary. In that case, Eq.~(\ref{eq:characteristic_equation_fievet}) is multiplied by $P_{\mathbf{Q}}^{-1}$ and after some rearrangements described in~\ref{appendix:T_eqn_inlet}, the system to solve for the unknowns wave amplitudes $\mathcal{N}_{1}^{*}$, $\mathcal{N}_{2}^{*}$, $\mathcal{N}_{3}^{*}$, $\mathcal{N}_{+}^{*}$, $\mathcal{N}_{5+1}^{*},\hdots,\mathcal{N}_{5+N_{s}}^{*}$ is given by:
\begin{align}
    |J|\frac{\partial u}{\partial t} &= n_{z}\left(\mathcal{N}_{2}^{*}+\mathcal{S}_{2}\right)-n_{y}\left(\mathcal{N}_{3}^{*}+\mathcal{S}_{3}\right) - \frac{n_{x}}{\sqrt{2}}\left(\mathcal{N}_{+}^{*}+\mathcal{S}_{+}-\mathcal{N}_{-}-\mathcal{S}_{-}\right) \label{eq:inlet_dudt}\\
    |J|\frac{\partial v}{\partial t} &= -n_{z}\left(\mathcal{N}_{1}^{*}+\mathcal{S}_{1}\right)+n_{x}\left(\mathcal{N}_{3}^{*}+\mathcal{S}_{3}\right) - \frac{n_{y}}{\sqrt{2}}\left(\mathcal{N}_{+}^{*}+\mathcal{S}_{+}-\mathcal{N}_{-}-\mathcal{S}_{-}\right) \label{eq:inlet_dvdt}\\
    |J|\frac{\partial w}{\partial t} &= n_{y}\left(\mathcal{N}_{1}^{*}+\mathcal{S}_{1}\right)-n_{x}\left(\mathcal{N}_{2}^{*}+\mathcal{S}_{2}\right) - \frac{n_{z}}{\sqrt{2}}\left(\mathcal{N}_{+}^{*}+\mathcal{S}_{+}-\mathcal{N}_{-}-\mathcal{S}_{-}\right) \label{eq:inlet_dwdt}\\
    \begin{split}
    |J|\frac{\partial T}{\partial t} &=\frac{T}{\rho}\left[n_{x}\left(\mathcal{N}_{1}^{*}+\mathcal{S}_{1}\right)+n_{y}\left(\mathcal{N}_{2}^{*}+\mathcal{S}_{2}\right)+n_{z}\left(\mathcal{N}_{3}^{*}+\mathcal{S}_{3}\right)\right]-\frac{T\left(\gamma -1\right)}{\sqrt{2}c}\left(\mathcal{N}_{+}^{*}+\mathcal{S}_{+}+\mathcal{N}_{-}+\mathcal{S}_{-}\right) \\
     &+ TW\sum\limits_{k=1}^{N_{s}}{\frac{\mathcal{N}_{5+k}^{*}+\mathcal{S}_{5+k}}{W_{k}}} \label{eq:inlet_dTdt}
     \end{split}
     \\
    |J|\frac{\partial Y_{k}}{\partial t} &=-\left(\mathcal{N}_{5+k}^{*}+\mathcal{S}_{5+k}\right)\hspace{0.20 cm}\text{for}\hspace{0.20 cm} k=1,N_{s}
    \label{eq:inlet_dYkdt}
\end{align}
Noting time-constant target values of the imposed primitive variables at the inlet boundary:
\begin{align}
    u = u^{t}\hspace{0.15 cm}\text{,}\hspace{0.15 cm}
    v = v^{t} \hspace{0.15 cm}\text{,}\hspace{0.15 cm}
    w = w^{t} \hspace{0.15 cm}\text{,}\hspace{0.15 cm}
    T = T^{t} \hspace{0.15 cm}\text{,}\hspace{0.15 cm}
    Y_{k} = Y_{k}^{t}\hspace{0.15 cm} \text{for}\hspace{0.15 cm} k=1,N_{s}
\end{align}
a relaxation procedure is applied as in the subsonic outflow condition, so that time derivatives in Eqs.~(\ref{eq:inlet_dudt}-\ref{eq:inlet_dYkdt}) are replaced by:
\begin{align}
\begin{split}
    \frac{\partial u}{\partial t} &= K_{\mathbf{u}}\left[u^{t}-u\right]\hspace{0.15 cm}\text{,}\hspace{0.15 cm}\frac{\partial v}{\partial t} = K_{\mathbf{u}}\left[v^{t}-v\right]\hspace{0.15 cm}\text{,}\hspace{0.15 cm}\frac{\partial w}{\partial t} = K_{\mathbf{u}}\left[w^{t}-w\right] \\
    \frac{\partial T}{\partial t} &= K_{T}\left[T^{t}-T\right]\hspace{0.15 cm}\text{,}\hspace{0.15 cm}\frac{\partial Y_{k}}{\partial t} = K_{Y_{k}}\left[Y_{k}^{t}-Y_{k}\right]
\end{split}
\end{align}
where $K_{\mathbf{u}}$, $K_{T}$ and $K_{Y_{k}}$ are respectively the relaxation rates for velocity components, temperature and species mass fractions. Note that here no condition on diffusive fluxes is imposed as stated in~\cite{poinsot1992boundary,sutherland2003improved}.

%% file: 4_WallBoundaryConditions.tex
\section{Symmetry and no-slip wall boundary conditions for SD\label{sec:wall_bc_SD}}
\noindent This section aims at defining what are the properties that symmetries and walls must satisfy and how to set them using SD formalism.

\subsection{General case}
\label{sub:general_case_symm_walls}
\noindent In this paragraph, a symmetry or a wall of normal $\mathbf{n}$ is considered.
\subsubsection{Symmetry boundary condition for a multispecies gas}
\noindent A symmetry for a multispecies gas is a boundary condition where the following conditions must be fulfilled:
\begin{align}
    u_{n}=0\hspace{0.1 cm}\text{,}\hspace{0.1 cm}
     \frac{\partial u_{t1}}{\partial n}=\frac{\partial u_{t2}}{\partial n}=0\hspace{0.1 cm}\text{and}\hspace{0.1 cm}
    \frac{\partial Y_{k}}{\partial n}=0\hspace{0.20 cm}\text{for}\hspace{0.20 cm} k=1,N_{s} \label{eq:symmetry_bc}
\end{align}
where $u_{t1/2}=\mathbf{u}.\mathbf{t}_{1/2}$ and $\left(\partial f/\partial n\right) \equiv \mathbf{\nabla}f.\mathbf{n} = n_{x}\left(\partial f/\partial x\right)+n_{y}\left(\partial f/\partial y\right)+n_{z}\left(\partial f/\partial z\right)$ is the notation for the gradient of any scalar function $f$ (for instance $u_{t1}$ in Eq.~(\ref{eq:symmetry_bc})) projected on $\mathbf{n}$. One condition remains to be found for pressure. Let's consider the three dimensional momentum equations taken from Eq.~(\ref{eq:momentum_cons_summary}) projected along $\mathbf{n}$:
\begin{align}
\begin{split}
    \frac{\partial \left(\rho u_{n}\right)}{\partial t} + \frac{\partial P}{\partial n} &+ \frac{\partial \left(\rho u u_{n}\right)}{\partial x} - \rho u^{2}\frac{\partial n_{x}}{\partial x} - \rho uv\frac{\partial n_{y}}{\partial x} - \rho uw\frac{\partial n_{z}}{\partial x} \\
    & + \frac{\partial \left(\rho v u_{n}\right)}{\partial y} - \rho uv\frac{\partial n_{x}}{\partial y} - \rho v^{2}\frac{\partial n_{y}}{\partial y} - \rho vw\frac{\partial n_{z}}{\partial y} \\
    & + \frac{\partial \left(\rho w u_{n}\right)}{\partial y} - \rho uw\frac{\partial n_{x}}{\partial z} - \rho vw\frac{\partial n_{y}}{\partial z} - \rho w^{2}\frac{\partial n_{z}}{\partial z} = n_{x}g_{x} + n_{y}g_{y} + n_{z}g_{z}
    \label{eq:normal_proj_momentum}
\end{split}
\end{align}
where $g_{x}$, $g_{y}$ and $g_{z}$ are respectively the viscous fluxes components of $x$, $y$ and $z$-momentum. Applying Eq.~(\ref{eq:symmetry_bc}), Eq.~(\ref{eq:normal_proj_momentum}) becomes:
\begin{align}
    \begin{split}
    \frac{\partial P}{\partial n} &= \rho u^{2}\frac{\partial n_{x}}{\partial x} + \rho uv\frac{\partial n_{y}}{\partial x} + \rho uw\frac{\partial n_{z}}{\partial x} 
     + \rho uv\frac{\partial n_{x}}{\partial y} + \rho v^{2}\frac{\partial n_{y}}{\partial y} + \rho vw\frac{\partial n_{z}}{\partial y}
    + \rho uw\frac{\partial n_{x}}{\partial z} + \rho vw\frac{\partial n_{y}}{\partial z} - \rho w^{2}\frac{\partial n_{z}}{\partial z} \\
    &+ n_{x}g_{x} + n_{y}g_{y} + n_{z}g_{z}
    \label{eq:dpdn_condition_symm}
    \end{split}
\end{align}
Eq.~(\ref{eq:dpdn_condition_symm}) shows that for a symmetry, the normal pressure gradient should balance the curvature and the projection of viscous fluxes on the normal to the symmetry plane. In practice, both the  curvature and viscous terms are neglected and the normal pressure gradient is set to zero at the symmetry.

\subsubsection{No-slip walls boundary conditions for a multispecies gas}
\noindent A no-slip wall for a multispecies gas is a boundary condition where the following conditions must be fulfilled: 
\begin{align}
    \mathbf{u}=\mathbf{0}\hspace{0.1 cm}\text{and}\hspace{0.1 cm} \frac{\partial Y_{k}}{\partial n}=0\hspace{0.20 cm}\text{for}\hspace{0.20 cm} k=1,N_{s} \label{eq:no_slip_wall_bc}
\end{align}
Consequently, Eq.~(\ref{eq:normal_proj_momentum}) becomes: 
\begin{align}
\frac{\partial P}{\partial n} &= n_{x}g_{x} + n_{y}g_{y} + n_{z}g_{z}\hspace{0.25 cm}
\label{eq:dpdn_condition_Navier}
\end{align}
Eq.~(\ref{eq:dpdn_condition_Navier}) shows that for no-slip walls, the normal pressure gradient should balance only the projection of viscous fluxes on the normal to the wall. Usually, these viscous fluxes are often neglected and the normal pressure gradient is also set to zero at no-slip walls. Then, two types of no-slip walls are commonly defined depending on the  temperature condition:
\begin{align}
    \frac{\partial T}{\partial n} &= 0\hspace{0.15 cm}\textbf{(Adiabatic no-slip wall)} \label{eq:cond_tempe_adia_wall}\\
    T &= T_{w}\hspace{0.15 cm}\textbf{(Isothermal no-slip wall)} \label{eq:cond_tempe_isoT_wall}
\end{align}
with $T_{w}$ a prescribed temperature to be set at the wall.

\subsection{How to satisfy a normal gradient condition $\left(\partial f/\partial n\right)$ in SD}
\noindent It was shown in paragraph \ref{sub:general_case_symm_walls} that for symmetry and no-slip walls, normal gradients to the boundary of some quantities as pressure, tangential velocities or mass fractions have to be imposed. Let's denote by $f$ any scalar function such as $P$, $T$ or $Y_{k}$ for which the normal gradient $\left(\partial f/\partial n\right)$ has to be imposed and is then assumed to be known here. The normal gradient along a normal $\mathbf{n}$ in the physical space is given by:
\begin{align}
    \begin{split}
    \frac{\partial f}{\partial n} &\equiv n_{x}\frac{\partial f}{\partial x}+n_{y}\frac{\partial f}{\partial y}+n_{z}\frac{\partial f}{\partial z} \\
    \frac{\partial f}{\partial n} &= n_{x}\left(\xi_{x}\frac{\partial f}{\partial \xi}+\eta_{x}\frac{\partial f}{\partial \eta}+\zeta_{x}\frac{\partial f}{\partial \zeta}\right) + n_{y}\left(\xi_{y}\frac{\partial f}{\partial \xi}+\eta_{y}\frac{\partial f}{\partial \eta}+\zeta_{y}\frac{\partial f}{\partial \zeta}\right) + n_{z}\left(\xi_{z}\frac{\partial f}{\partial \xi}+\eta_{z}\frac{\partial f}{\partial \eta}+\zeta_{z}\frac{\partial f}{\partial \zeta}\right) \\
    \frac{\partial f}{\partial n} &= \left(n_{x}\xi_{x}+n_{y}\xi_{y}+n_{z}\xi_{z}\right)\frac{\partial f}{\partial \xi} + \left(n_{x}\eta_{x}+n_{y}\eta_{y}+n_{z}\eta_{z}\right)\frac{\partial f}{\partial \eta} +\left(n_{x}\zeta_{x}+n_{y}\zeta_{y}+n_{z}\zeta_{z}\right)\frac{\partial f}{\partial \zeta} 
    \end{split} \label{eq:df_dn_1}
\end{align}
Taking for example a $\xi$-normal boundary, $\mathbf{n}$ will be given by the following relation~\cite{veilleux2021extension}:
\begin{align}
    \mathbf{n} \equiv \left(n_{x},n_{y},n_{z}\right)^{\mathrm{T}} = |J|\left(J^{-1}\right)^{\mathrm{T}}.\mathbf{n}_{iso} = |J|\left(\xi_{x},\xi_{y},\xi_{z}\right)^{\mathrm{T}}
\end{align}
where $\mathbf{n}_{iso}=\left(1,0,0\right)^{\mathrm{T}}$ for a $\xi$-normal boundary. Consequently, Eq.~(\ref{eq:df_dn_1}) is now:
\begin{equation}
    \frac{\partial f}{\partial n} = \frac{\left(n_{x}^{2}+n_{y}^{2}+n_{z}^{2}\right)}{|J|}\frac{\partial f}{\partial \xi} + \left(n_{x}\eta_{x}+n_{y}\eta_{y}+n_{z}\eta_{z}\right)\frac{\partial f}{\partial \eta} +\left(n_{x}\zeta_{x}+n_{y}\zeta_{y}+n_{z}\zeta_{z}\right)\frac{\partial f}{\partial \zeta} \label{eq:df_dn_2}
\end{equation}
Finally, by denoting $A_{FP}=\sqrt{n_{x}^{2}+n_{y}^{2}+n_{z}^{2}}$, Eq.~(\ref{eq:df_dn_1}) can be expressed for $\left(\partial f/\partial \xi\right)$:
\begin{equation}
    \frac{\partial f}{\partial \xi} = \frac{|J|}{A_{FP}^{2}}\left(\frac{\partial f}{\partial n} - \left(n_{x}\eta_{x}+n_{y}\eta_{y}+n_{z}\eta_{z}\right)\frac{\partial f}{\partial \eta} - \left(n_{x}\zeta_{x}+n_{y}\zeta_{y}+n_{z}\zeta_{z}\right)\frac{\partial f}{\partial \zeta}\right)
    \label{eq:compute_dfdxi}
\end{equation}
Since $\left(\partial f/\partial n\right)$, the metrics and the derivatives in tangential directions of the boundary are known, the RHS of Eq.~(\ref{eq:compute_dfdxi}) can be computed. Then, $\left(\partial f/\partial \xi\right)$ is known and the methodology explained in paragraph \ref{sub:useful_formula_nscbc} is used to compute the value of $f$ at the boundary FP.

\subsection{Application in the SD context}
\noindent In a SD context, a boundary condition is a particular kind of interface, noted $I$, where only one state, the one from the computational domain, is known. The remaining state is outside the computational domain and has to be determined to satisfy the type of boundary condition. As mentioned in paragraph~\ref{sub:use_prim_var_FP}, at FP primitive variables are employed. Assuming that the boundary \textit{left state} $\mathbf{Q}_{FP}^{L}$ is the known state from the interior domain, the boundary \textit{right state} $\mathbf{Q}_{FP}^{R}$ must be determined to fulfill the symmetry or adiabatic or isothermal no-slip wall boundary conditions. This methodology for applying boundary condition in a compact high-order context is called the \textit{Weak-Riemann} approach by Mengaldo et al.~\cite{mengaldo2014guide}. In their paper, the authors explained the implementation of boundary conditions in a DG and FR context for the compressible Navier-Stokes equations and a monospecies gas. One major difference with the SD method is that DG and FR use a single set of points whereas SD use SP for the interior domain discretization and FP for the boundaries. As SP are strictly inside the computational domain, their velocity is non-zero and the extrapolated velocity from SP to a boundary FP (the \textit{left} velocity at the boundary) will never be zero. On the contrary, for DG or FR, this \textit{left} velocity is directly imposed to zero and used in the numerical scheme. Let's now consider a boundary located at $\xi=1$ as in Figure~\ref{fig:boundary_general_case} with its associated convective and diffusive fluxes respectively $\widehat{\mathbf{E}}_{c}^{I}$ and $\widehat{\mathbf{E}}_{d}^{I}$.

\subsubsection{Symmetry boundary condition}
\noindent From Eq.~(\ref{eq:symmetry_bc}) completed by Eq.~(\ref{eq:dpdn_condition_symm}), the right state is set as:
\begin{align}
    \mathbf{Q}_{FP}^{R} = \left(
    \begin{array}{c}
         T_{FP}^{R}  \\
         u_{n,FP}^{R} \\
         u_{t1,FP}^{R} \\
         u_{t2,FP}^{R} \\
         P_{FP}^{R} \\
         Y_{k,FP}^{R} 
    \end{array}
    \right) = \left(
    \begin{array}{c}
         T_{p+2,FP}  \\
         0 \\
         u_{t1,p+2,FP} \\
         u_{t2,p+2,FP} \\
         P_{p+2,FP} \\
         Y_{k,p+2,FP}
    \end{array}
    \right) 
    \label{eq:right_state_symm}
\end{align}
where values with subscript $\left(p+2,FP\right)$ have been computed using Eq.~(\ref{eq:get_fpPLUS2_hat_FP}) where each value of $\left(\partial f/\partial \xi\right)$ was found using Eq.~(\ref{eq:compute_dfdxi}) for each value of $\left(\partial f/\partial n\right)$ corresponding to the symmetry case. In Eq.~(\ref{eq:right_state_symm}), velocity components are expressed in the $\left(\mathbf{n},\mathbf{t}_{1},\mathbf{t}_{2}\right)$ basis. Thus, to retrieve the cartesian velocity at $\left(p+2,FP\right)$ the transformation shown in Eq.~(\ref{eq:link_local2cartesian_vel}) is applied:
\begin{align}
    \begin{split}
        u_{FP}^{R} &= u_{t1,FP}^{R}t_{1x} + u_{t2,FP}^{R}t_{2x} \\
        v_{FP}^{R} &= u_{t1,FP}^{R}t_{1y} + u_{t2,FP}^{R}t_{2y} \\
        w_{FP}^{R} &= u_{t1,FP}^{R}t_{1z} + u_{t2,FP}^{R}t_{2z}
    \end{split}
    \label{eq:link_local2cartesian_vel}
\end{align}
Once left and right states are known, the convective and diffusive fluxes at interface $I$ are computed as follows:
\begin{align}
    \widehat{\mathbf{E}}_{c}^{I} = \widehat{\mathbf{E}}_{c}^{I}\left(\mathbf{Q}_{FP}^{L},\mathbf{Q}_{FP}^{R}\right)\hspace{0.10 cm}\text{and}\hspace{0.10 cm}
    \widehat{\mathbf{E}}_{d}^{I} = \widehat{\mathbf{E}}_{d}^{I}\left(\mathbf{Q}_{FP}^{R},\left(\nabla \mathbf{Q}\right)_{FP}^{R}\right) \label{eq:convective_and_diffusive_flux_bnd}
\end{align}
with $\left(\nabla \mathbf{Q}\right)_{FP}^{R}$ the gradient of each variable in $\mathbf{Q}_{FP}$ computed using the value of $ \mathbf{Q}_{FP}^{R}$ on the symmetry. 

\subsubsection{Adiabatic no-slip wall}
\label{subsub:adia_wall}
\noindent From Eqs.~(\ref{eq:no_slip_wall_bc}-\ref{eq:dpdn_condition_Navier}) completed by Eq.~(\ref{eq:cond_tempe_adia_wall}), the right state is set as:
\begin{align}
    \mathbf{Q}_{FP}^{R} = \left(
    \begin{array}{c}
         T_{FP}^{R}  \\
         u_{FP}^{R} \\
         v_{FP}^{R} \\
         w_{FP}^{R} \\
         P_{FP}^{R} \\
         Y_{k,FP}^{R} 
    \end{array}
    \right) = \left(
    \begin{array}{c}
         T_{p+2,FP}  \\
         0 \\
         0 \\
         0 \\
         P_{p+2,FP} \\
         Y_{k,p+2,FP}
    \end{array}
    \right) 
    \label{eq:right_state_adia_wall}
\end{align}
$\widehat{\mathbf{E}}_{c}^{I}$ and $\widehat{\mathbf{E}}_{d}^{I}$ are also computed using Eq.~(\ref{eq:convective_and_diffusive_flux_bnd}). Another choice for the convective fluxes suggested in~\cite{mengaldo2014guide} as the \textit{Weak-Riemann-A1} approach, would be to link the left and right states in order to have an intermediate Riemann state where the normal component of velocity is zero which is not the case if Eq.~(\ref{eq:right_state_adia_wall}) is used in Eq.~(\ref{eq:convective_and_diffusive_flux_bnd}). Both approaches gave the same results for the cases considered in this work but the one of Eq.~(\ref{eq:right_state_adia_wall}) is prefered here since the pressure at the wall really follows Eq.~(\ref{eq:dpdn_condition_Navier}).

\subsubsection{Isothermal no-slip wall}
\label{subsub:isoT_wall}
\noindent From Eqs.~(\ref{eq:no_slip_wall_bc}-\ref{eq:dpdn_condition_Navier}) but this time completed by Eq.~(\ref{eq:cond_tempe_isoT_wall}), the right state is set as:
\begin{align}
    \mathbf{Q}_{FP}^{R} = \left(
    \begin{array}{c}
         T_{FP}^{R}  \\
         u_{FP}^{R} \\
         v_{FP}^{R} \\
         w_{FP}^{R} \\
         P_{FP}^{R} \\
         Y_{k,FP}^{R} 
    \end{array}
    \right) = \left(
    \begin{array}{c}
         T_{w}  \\
         0 \\
         0 \\
         0 \\
         P_{p+2,FP} \\
         Y_{k,p+2,FP}
    \end{array}
    \right) 
    \label{eq:right_state_isoT_wall}
\end{align}
with again $\widehat{\mathbf{E}}_{c}^{I}$ and $\widehat{\mathbf{E}}_{d}^{I}$ obtained using Eq.~(\ref{eq:convective_and_diffusive_flux_bnd}). The same remarks on the normal component of velocity as in paragraph \ref{subsub:adia_wall} also hold here.

%% file: 8_ValidationTestCases.tex
\section{Validation test cases}
\label{sec:test_cases}
\noindent To validate the implementation of reacting flow models in JAGUAR, 1D and 2D simulations of laminar flames with different chemistries have been performed. For all the simulations, JAGUAR results are compared with results obtained with the reference solver AVBP~\cite{schonfeld1999steady,gourdain2009high} developed by CERFACS. AVBP solves exactly the same reacting NSE for a multispecies gas with the same transport model. All AVBP simulations were done using the TTGC scheme~\cite{colin2000development} for convective fluxes without artificial viscosity for a fair comparison with JAGUAR. For diffusion fluxes, a finite element scheme of order 2 was used~\cite{colin2000development}. One-dimensional cases are also compared to the reference chemistry code CANTERA~\cite{cantera}.

\subsection{One-dimensional flame using a two-reactions chemistry}
\label{sub:1D_BFER_DNS}
\begin{figure}[!h]
    \centering
    \includegraphics[width=\Lfig, height=\Hfig]{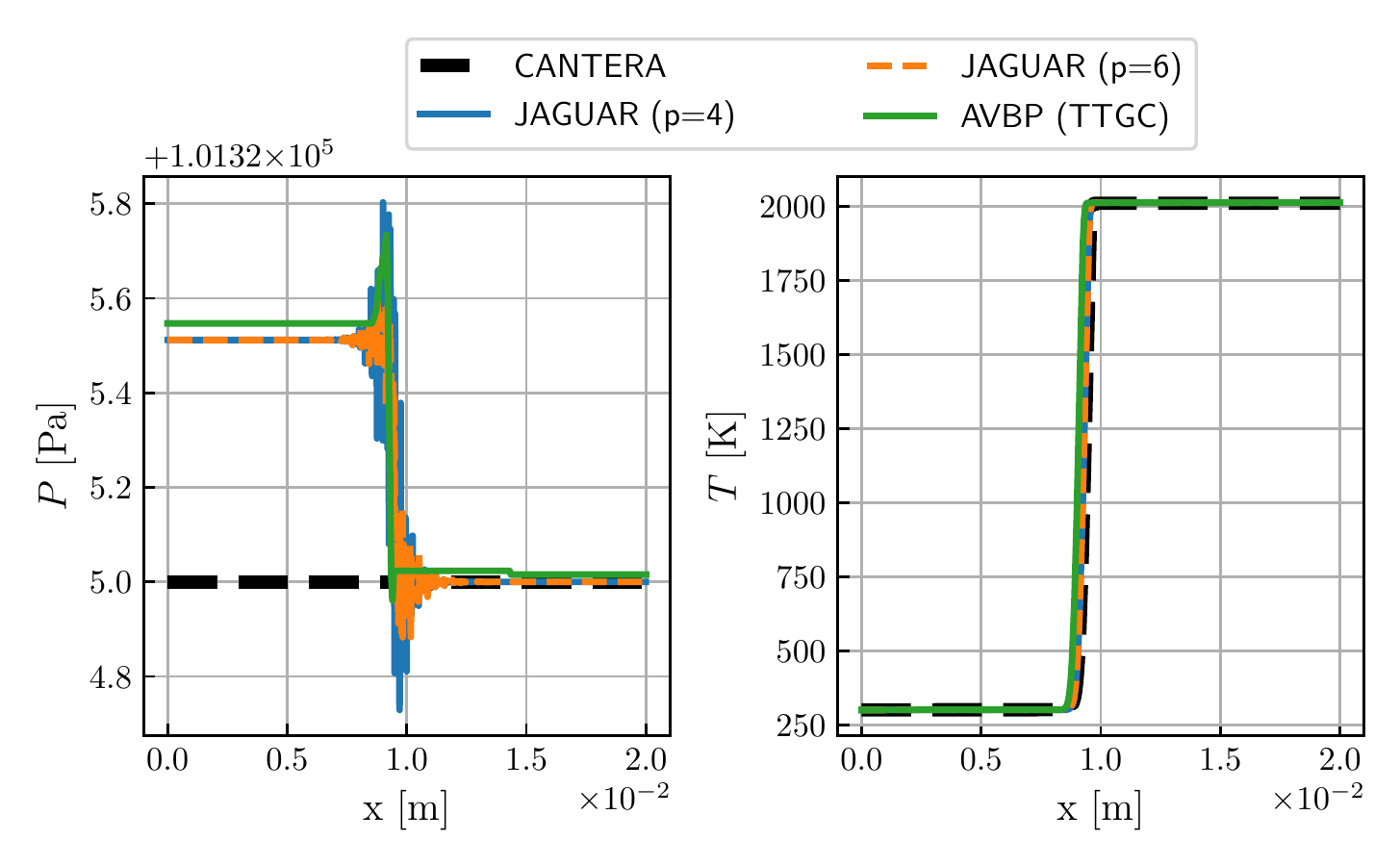}
    \caption{Comparison of pressure and temperature profiles between CANTERA, AVBP and JAGUAR for a 1D premixed methane-air flame using the CH4/Air-2S-BFER chemical scheme at $\phi=0.8$.}
    \label{fig:compa_NS_BFER_1D_DNS}
\end{figure}

\begin{figure}[!h]
    \centering
    \includegraphics[width=\Lfig, height=\Hfig]{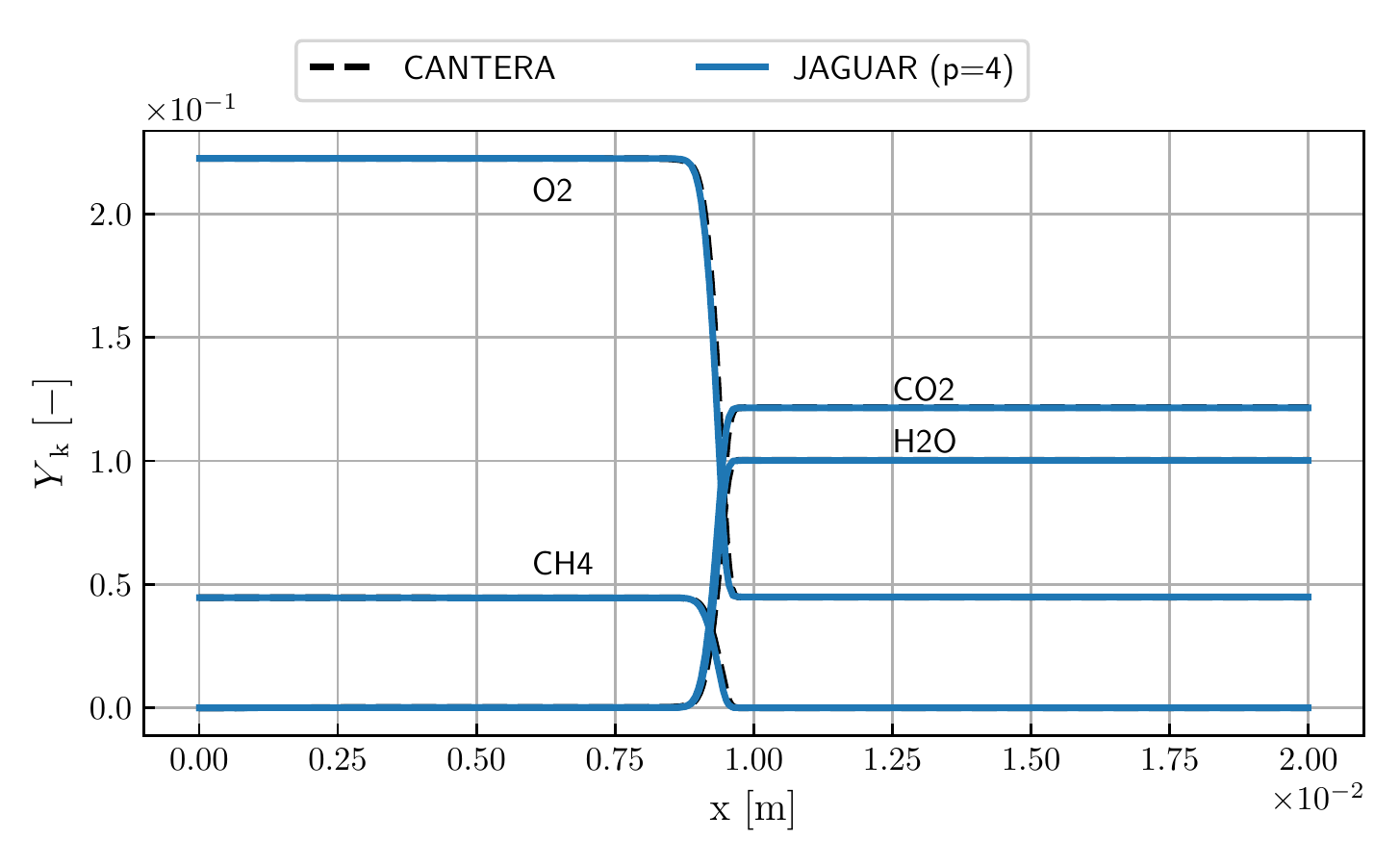}
    \caption{Comparison of species mass fractions profiles between CANTERA and JAGUAR at $p=4$ for a 1D methane-air flame using the CH4/Air-2S-BFER chemical scheme at $\phi=0.8$.}
    \label{fig:compa_spec_BFER_1D_DNS}
\end{figure}

\noindent The objectives of this test case are to validate in JAGUAR:
\begin{itemize}
    \item[$\bullet$] The source term and species transport computations with a reduced two-reactions scheme.
    \item[$\bullet$] The implementation of JANAF thermochemical tables.
    \item[$\bullet$] The NSCBC inlet and outlet boundary conditions on a 1D reacting case.
\end{itemize}
A one-dimensional methane-air premixed flame is considered. The chemical scheme is the two-steps CH4/Air-2S-BFER developed in~\cite{franzelli2010two}. The characteristics of this flame are given in Table \ref{tab:1D_BFER_phi08_charac} where $\phi$ is the global equivalence ratio, $T^{f}$ is the fresh gases temperature, $T^{b}$ and $S_{L}^{0}$ are respecticely the burnt gases temperature and the laminar flame speed estimated by CANTERA and $\delta_{L}^{0}$ is the laminar flame thickness also obtained from the CANTERA solution using~\cite{poinsot2005theoretical}:
\begin{equation}
    \delta_{L}^{0} = \frac{T^{b}-T^{f}}{\max\left(|\frac{\partial T}{\partial x}|\right)}
    \label{eq:Def_deltaL_gradT}
\end{equation}
\begin{table}[!h]
    \centering
    \begin{tabular}{|c|c|c|c|c|c|}
    \hline
        $\phi$ [-] &  $T^{f}\ [\mathrm{K}]$ & $T^{b}\ [\mathrm{K}]$ & $P\ [\mathrm{Pa}]$ & $\delta_{L}^{0}$ [m] & $S_{L}^{0}\ [\mathrm{cm.s^{-1}}]$\\
        \hline
        0.8 & 300 & 2010 & 101325 & $4.30\times 10^{-4}$ & 28.15 \\
        \hline
    \end{tabular}
    \caption{Characteristics of the 1D methane-air premixed flame for the CH4/Air-2S-BFER scheme. $T^{b}$ and $S_{L}^{0}$ are the values given by CANTERA.}
    \label{tab:1D_BFER_phi08_charac}
\end{table}
\noindent The computational domain is a 1D segment of length $L_{x}=0.02$ m discretized with $N_{e}$ elements. The left boundary condition is a NSCBC subsonic inflow imposing $u$, $v$, $w$, $T$ and $Y_{k}$ and the right boundary is a NSCBC subsonic outflow imposing a static pressure. To allow a fair comparison between JAGUAR and AVBP, all calculations must be done with similar number of points in the 1D domain. These points are called degree of freedom (DOFs): they correspond to cell nodes for AVBP and to SP for JAGUAR. In space dimension $d$, a standard hexahedral element where $\widehat{\mathbf{U}}$ varies as a $p$-degree polynomial contains $N_{SP}=\left(p+1\right)^{d}$ SP so that if all these elements have the same degree $p$, the number of DOFs inside the computational domain (total number of SP) is:
\begin{align}
    \mathrm{DOFs^{SD}} = N_{e}\times N_{SP} = N_{e}\left(p+1\right)^{d}
    \label{eq:def_DOFs_SD}
\end{align}
Here it was chosen to keep the number of DOFs around 400 in order to have a number of points $n_{f}$ in the flame region of about:
\begin{align}
    n_{f} = \frac{\delta_{L}^{0}}{\Delta x} = \frac{4.30\times 10^{-4}}{\frac{0.02}{400}} \approx 9 
\end{align}
which is sufficient to well-resolve the flame front. While this imposes a fixed number of 400 nodes in AVBP, JAGUAR has the possibility to use different values of $N_{e}$ depending on the polynomial order $p$ to keep DOFs$^\mathrm{SD}$ close to 400 as summed up in Table~\ref{tab:Ne_with_order_1D_flame}.
\begin{table}[!h]
    \centering
    \begin{tabular}{|c|c|c|c|c|}
        \hline
       $p$ & 3 & 4 & 5 & 6 \\
       \hline
       $N_{e}$ & 100 & 80 & 67 & 57 \\
       \hline
    \end{tabular}
    \caption{Values of $N_{e}$ associated to each polynomial degree $p$ to keep DOFs$^\mathrm{SD}$ around 400 in the 1D domain.}
    \label{tab:Ne_with_order_1D_flame}
\end{table}
Both JAGUAR and AVBP simulations are initialized using a CANTERA solution which contains approximately 30 points in the flame region. After a transient phase due to the transition from an incompressible to a compressible solution, the solution converges and final profiles of pressure and temperature are represented in Figure~\ref{fig:compa_NS_BFER_1D_DNS} for CANTERA, AVBP and JAGUAR with $p=4$ and $p=6$. Major mass fractions are plotted in Figure~\ref{fig:compa_spec_BFER_1D_DNS} for CANTERA and JAGUAR at $p=4$ only since AVBP and JAGUAR at $p=6$ gave the same results.

\noindent All profiles are in excellent agreement. The pressure jump through the flame front is captured by both JAGUAR and AVBP while CANTERA runs at constant pressure. Theoretically, this pressure drop is given by~\cite{poinsot2005theoretical}:
\begin{equation}
    P^{b} - P^{f} = \rho^{f}\left(S_{L}^{0}\right)^{2}\left(1-\frac{T^{b}}{T^{f}}\right)
    \label{eq:PressureDrop_1D_Flame}
\end{equation}
where $P^{b}$ and $P^{f}$ are respectively the pressure in burnt and fresh gases and $\rho^{f}$ is the fresh gases density. Using values of Table.~\ref{tab:1D_BFER_phi08_charac}, Eq.~(\ref{eq:PressureDrop_1D_Flame}) leads to a pressure drop of $-0.511$ Pa i.e., very close to the value of $-0.512$ Pa measured from JAGUAR and AVBP solutions. Finally, flame speeds can be compared. They are computed using the consumption speed:
\begin{align}
    S_{c} = \frac{-1}{\rho^{f}Y_{F}^{f}-\rho^{b}Y_{F}^{b}}\int_{-\infty}^{+\infty}{\dot{\omega}_{F}\; d\mathbf{n}}
    \label{eq:Def_consumption_speed}
\end{align}
since for a steady laminar unstretched flame: $S_{L}^{0}=S_{c}$. In Eq.~(\ref{eq:Def_consumption_speed}), $\mathbf{n}$ is the local normal to the flame front and $Y_{F}^{f}$ and $Y_{F}^{b}$ respectively the fuel mass fractions in fresh and burnt gases. Flame speeds obtained with AVBP and JAGUAR solutions are summed up in Table~\ref{tab:compa_SL_BFER_phi08} where $\epsilon_{rel}$ is the relative error compared to CANTERA reference value given in Table~\ref{tab:1D_BFER_phi08_charac}.
\begin{table}[!h]
    \centering
    \begin{tabular}{|c|c|c|c|c|c|}
    \hline
        Code & AVBP & JAGUAR (p=3) & JAGUAR (p=4) & JAGUAR (p=5) & JAGUAR (p=6)\\
    \hline
       $S_{L}^{0}\ [\mathrm{cm.s^{-1}}]$  & 28.31 & 28.17 & 28.16 & 28.16 & 28.15 \\
    \hline
    $\epsilon_{rel}$ in $\%$ & 0.568 & 0.071 & 0.036 & 0.036 & 0.0 \\
    \hline
    \end{tabular}
    \caption{Comparison of flame speed obtained with AVBP and JAGUAR at different orders for a 1D premixed methane-air flame using the CH4/Air-2S-BFER chemical scheme at $\phi=0.8$.}
    \label{tab:compa_SL_BFER_phi08}
\end{table}
\noindent Flame speeds estimated by JAGUAR are in very good agreement with the value given by CANTERA which shows its capability to well capture the flame front.

\subsection{One-dimensional flame using Analytically Reduced Chemistry}
\label{sub:1D_ARC_DNS}
\noindent The objective of this test case is to simulate a one-dimensional flame with JAGUAR using a more detailed chemistry including more species. Recently, new types of chemical schemes have been introduced as intermediate mechanisms between detailed ones with hundreds of species and global ones with a maximum of ten species. The extremely high number of species of detailed mechanisms makes them far too costly for both academic and industrial 3D configurations. Conversely global schemes, although much less costly and easily affordable, do not reproduce chemical pathways and pollutant formation processes. An alternative is the so-called Analytically Reduced Chemistry (ARC), reducing detailed mechanism with physically-oriented methods, to between 10 to 30 species which is today affordable in CFD simulations. Mechanisms derived with the ARC methodology have the particularity to rely directly upon the detailed mechanism using the true parameters in Arrhenius's law. Therefore, ARC mechanisms are expected to be quite accurate with a computational cost still reasonable for today computers. The chemical scheme considered here is a methane-air mechanism derived with software ARCANE~\cite{cazeres2021fully} from the GRI-3.0~\cite{SiteGRI30} detailed mechanism. It is composed of 16 transported species, 250 chemical reactions and 10 species in quasi-steady-state (QSS)~\cite{lu2006systematic}. These QSS species have a very small characteristic timescale and are considered to have a zero net chemical source term so that their concentration is computed as a function of the concentrations of the other species. Therefore, they are not computed with a transport equation which reduces the computational time. 

\begin{figure}[!h]
    \centering
    \includegraphics[width=\Lfig, height=\Hfig]{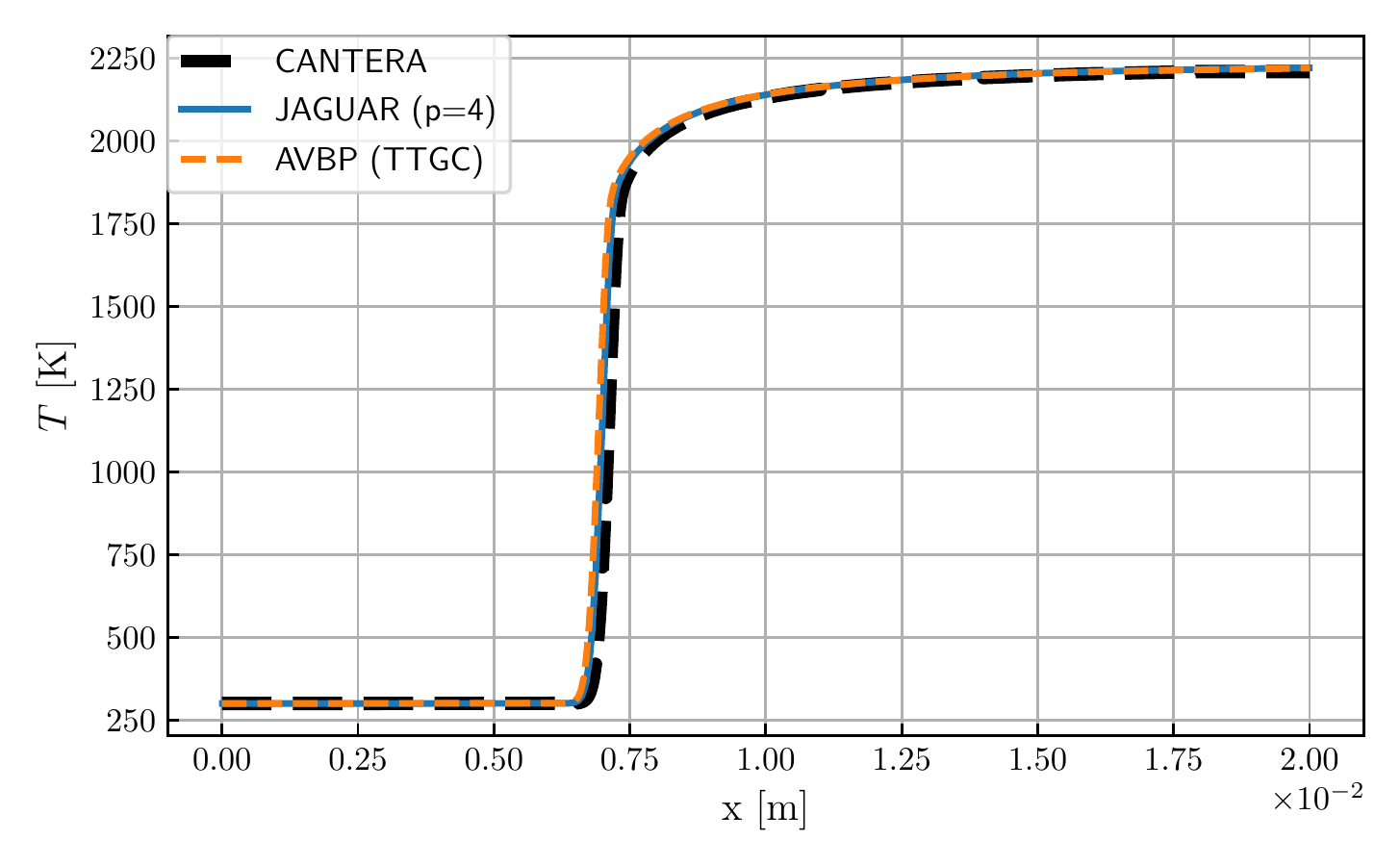}
    \caption{Comparison of temperature profiles between CANTERA, AVBP and JAGUAR for a 1D premixed methane-air flame using an ARC chemical scheme at $\phi=1.0$.}
    \label{fig:compa_NS_ARC_1D_DNS}
\end{figure}

\begin{figure}[!h]
    \centering
    \includegraphics[width=\Lfig, height=\Hfig]{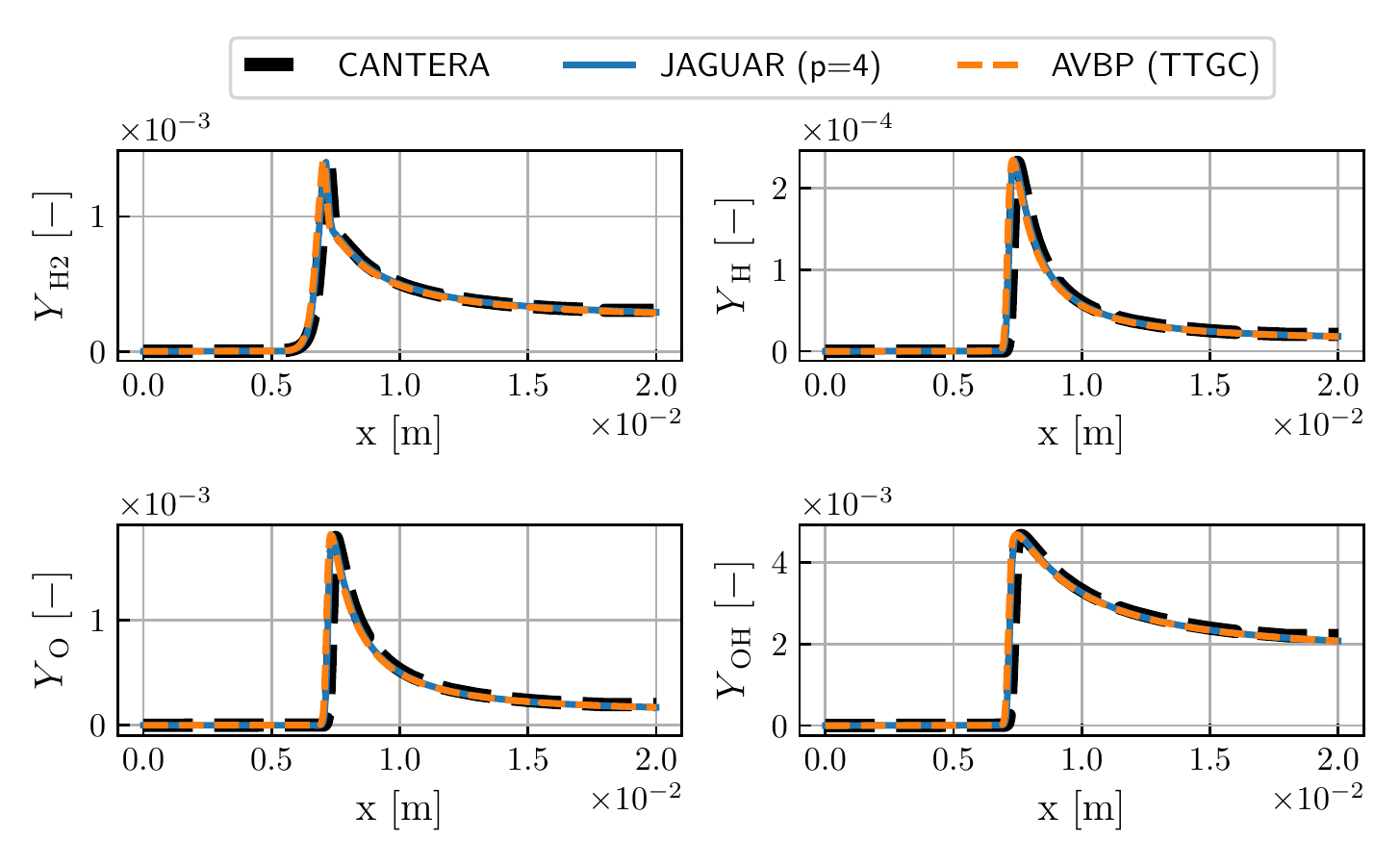}
    \caption{Comparison of species mass fractions profiles between CANTERA, AVBP and JAGUAR for a 1D premixed methane-air flame using an ARC chemical scheme at $\phi=1.0$.}
    \label{fig:compa_YH2_YH_YO_YOH_ARC_1D_DNS}
\end{figure}

\begin{figure}[!h]
    \centering
    \includegraphics[width=\Lfig, height=\Hfig]{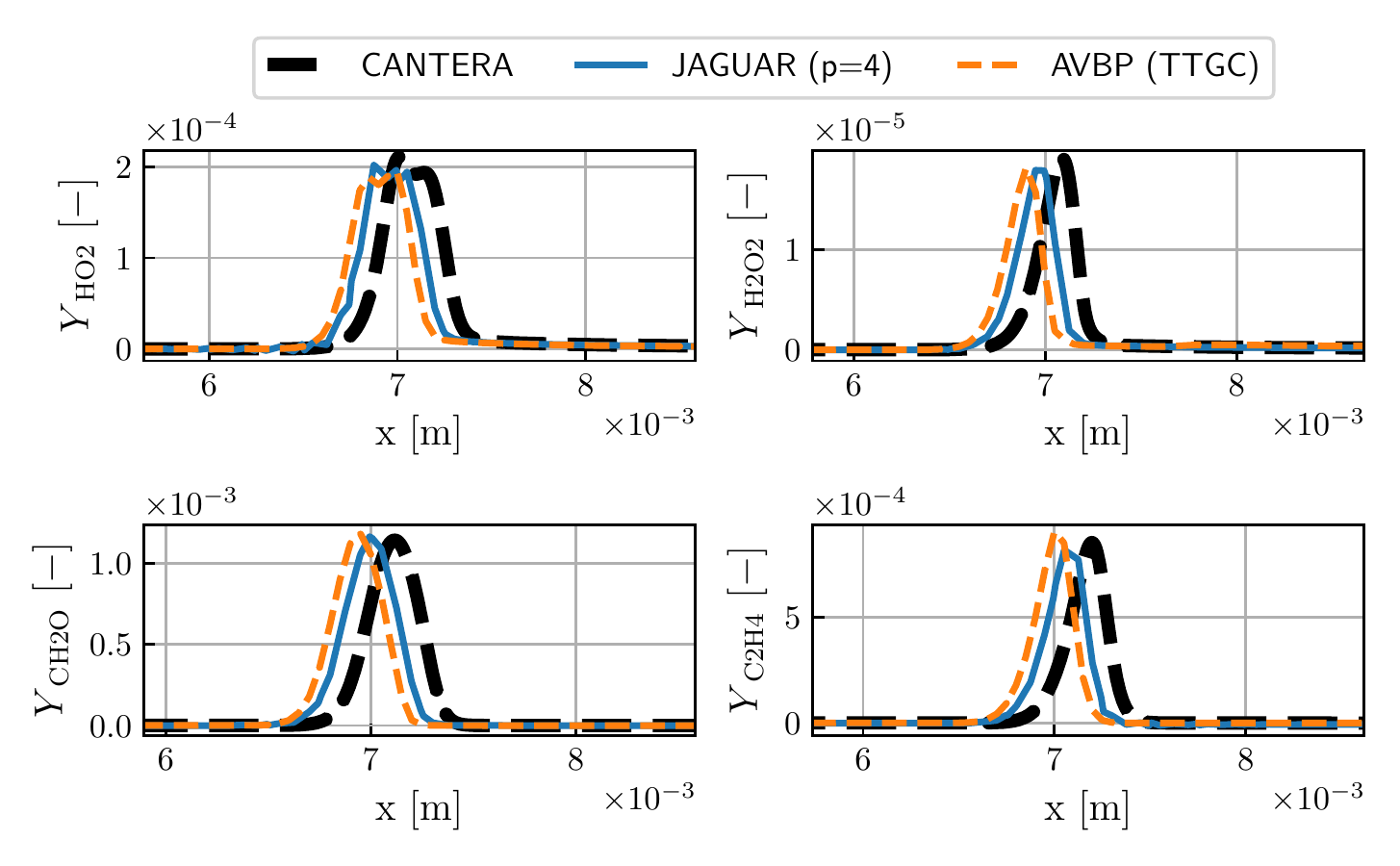}
    \caption{Comparison of species mass fractions profiles between CANTERA, AVBP and JAGUAR for a 1D premixed methane-air flame using an ARC chemical scheme at $\phi=1.0$.}
    \label{fig:compa_YHO2_YH2O2_YCH2O_YC2H4_ARC_1D_DNS}
\end{figure}

\noindent The characteristics of this flame are given in Table \ref{tab:1D_ARC_phi1_charac}.
\begin{table}[!h]
    \centering
    \begin{tabular}{|c|c|c|c|c|c|}
    \hline
        $\phi$ [-] &  $T^{f}\ [\mathrm{K}]$ & $T^{b}\ [\mathrm{K}]$ & $P\ [\mathrm{Pa}]$ & $\delta_{L}^{0}$ [m] & $S_{L}^{0}\ [\mathrm{cm.s^{-1}}]$\\
        \hline
        1.0 & 300 & 2210 & 100000 & $4.39\times 10^{-4}$ & 37.86 \\
        \hline
    \end{tabular}
    \caption{Characteristics of the 1D methane-air premixed flame using an ARC mechanism. $T^{b}$ and $S_{L}^{0}$ are the values given by CANTERA.}
    \label{tab:1D_ARC_phi1_charac}
\end{table}
\noindent The flame thickness is almost the same as the one obtained with the CH4/Air-2S-BFER at $\phi=0.8$ already simulated in paragraph~\ref{sub:1D_BFER_DNS}. Then, the computational domain is the same 1D segment of length $L_{x}=0.02$ m discretized here only for the case with $N_{e}=80$ elements and a polynomial order sets to $p=4$. This should be sufficient to well-resolve the flame front even if the chemistry is stiffer in that case. Temperature profiles are represented in Figure~\ref{fig:compa_NS_ARC_1D_DNS} and some species mass fractions profiles are represented in Figures~\ref{fig:compa_YH2_YH_YO_YOH_ARC_1D_DNS} and~\ref{fig:compa_YHO2_YH2O2_YCH2O_YC2H4_ARC_1D_DNS}.

\noindent JAGUAR results are very close to AVBP and CANTERA results which shows its capability to simulate more complex chemistry. Moreover, the flame speeds predicted by JAGUAR (38.06 cm.$\mathrm{s}^{-1}$) and AVBP (38.09 cm.$\mathrm{s}^{-1}$) are almost identical meaning that JAGUAR gives comparable results to a well established combustion solver.

\subsection{Two-dimensional flame using a two-reactions chemistry}
\begin{figure}[!h]
    \centering
    \includegraphics{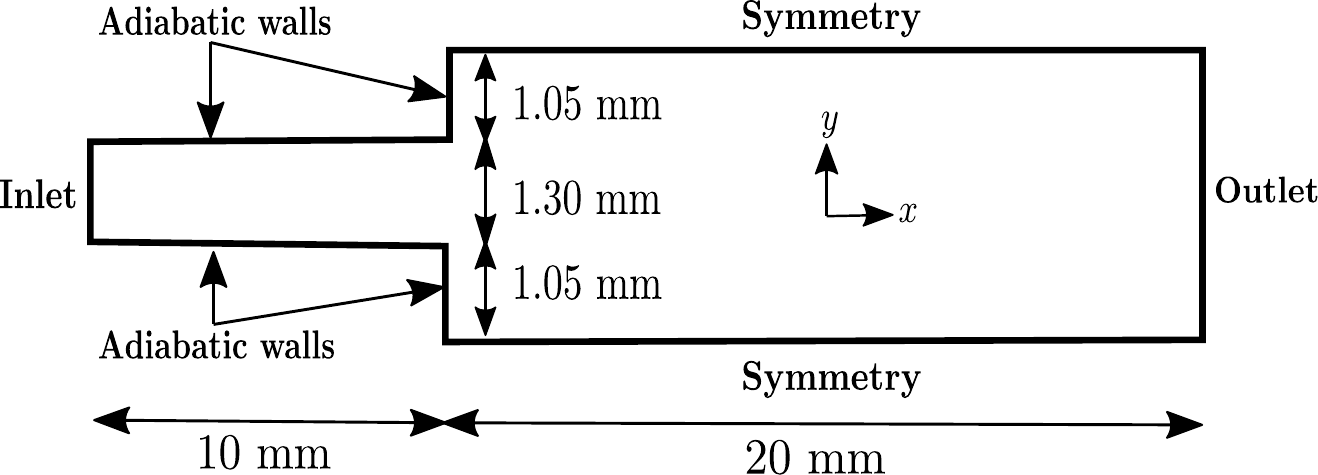}
    \caption{Computational domain and boundary conditions for the 2D burner case.}
    \label{fig:domain_2D_burner}
\end{figure}

\noindent To go a step further in the validation, a 2D methane-air premixed flame with walls and symmetry boundary conditions is now computed. The geometry is the one of a 2D burner described in Figure~\ref{fig:domain_2D_burner}. 
\noindent Fresh gases enter the burner axially (x-axis) through a NSCBC subsonic inflow imposing a parabolic profile given by Eq.~(\ref{eq:parabolic_profile_2D_burner}):
\begin{align}
    u\left(y\right)=u_{0}\left(1-\frac{y^{2}}{l_{0}^{2}}\right)
    \label{eq:parabolic_profile_2D_burner}
\end{align}
where $u_{0}=4\ \mathrm{m.s^{-1}}$ and $l_{0}=0.65$ mm. Equivalence ratio and fresh gas temperature are also set respectively to 0.8 and 300 K at the inlet. From $x=0$ to $x=10$ mm side walls are adiabatic and from $x=10$ mm to $x=30$ mm symmetry boundary conditions are applied. At the outlet, a subsonic outflow imposing $P=101325$ Pa is employed. The two-steps mechanism is still the CH4/Air-2S-BFER. For the JAGUAR simulation, the domain is discretized using 1216 quadrilateral elements with a polynomial of degree $p=4$ (order 5). Consequently, the total number of DOFs inside the SD mesh is 30400. AVBP solution domain is still discretized with almost the same number of DOFs (31574 nodes in that case) to have a fair comparison. 

\begin{figure}[!h]
    \centering
    \includegraphics[width=12cm, height=5.5cm]{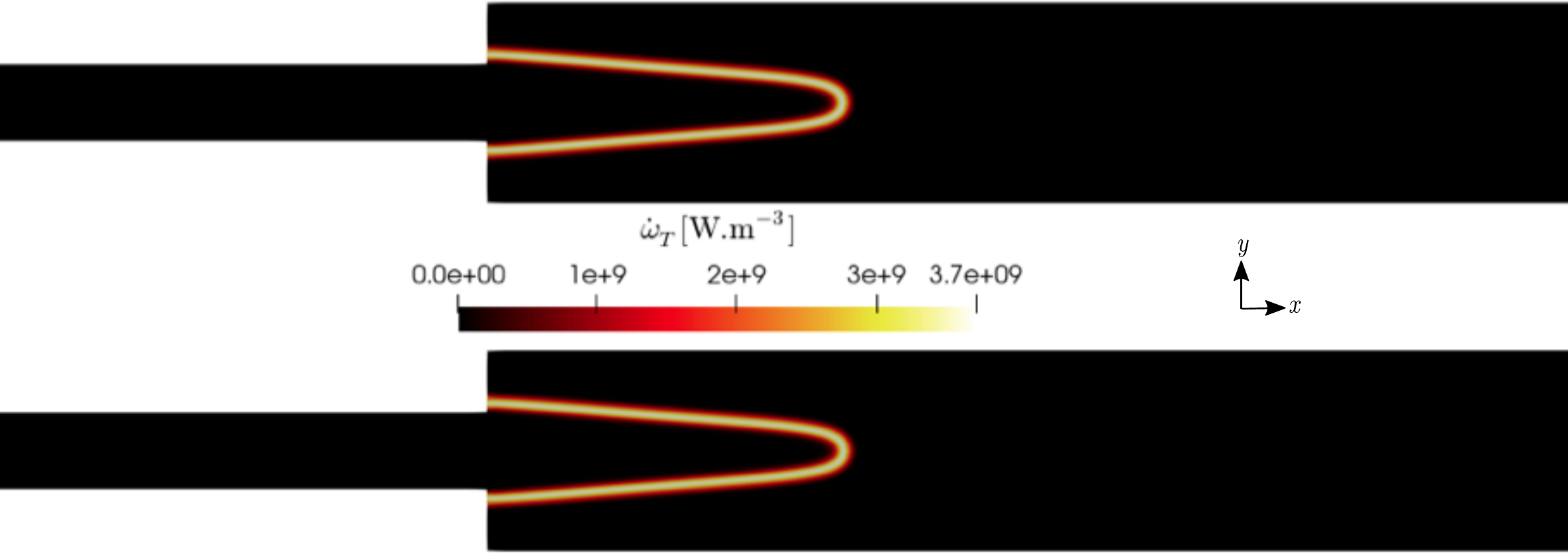}
    \caption{Comparison between JAGUAR (top) and AVBP (bottom) on the 2D heat release rate field for the 2D burner case.}
    \label{fig:compa_hr_top_JAG_bot_AVBP}
\end{figure}

\begin{figure}[!h]
    \centering
    \includegraphics[width=\Lfig, height=\Hfig]{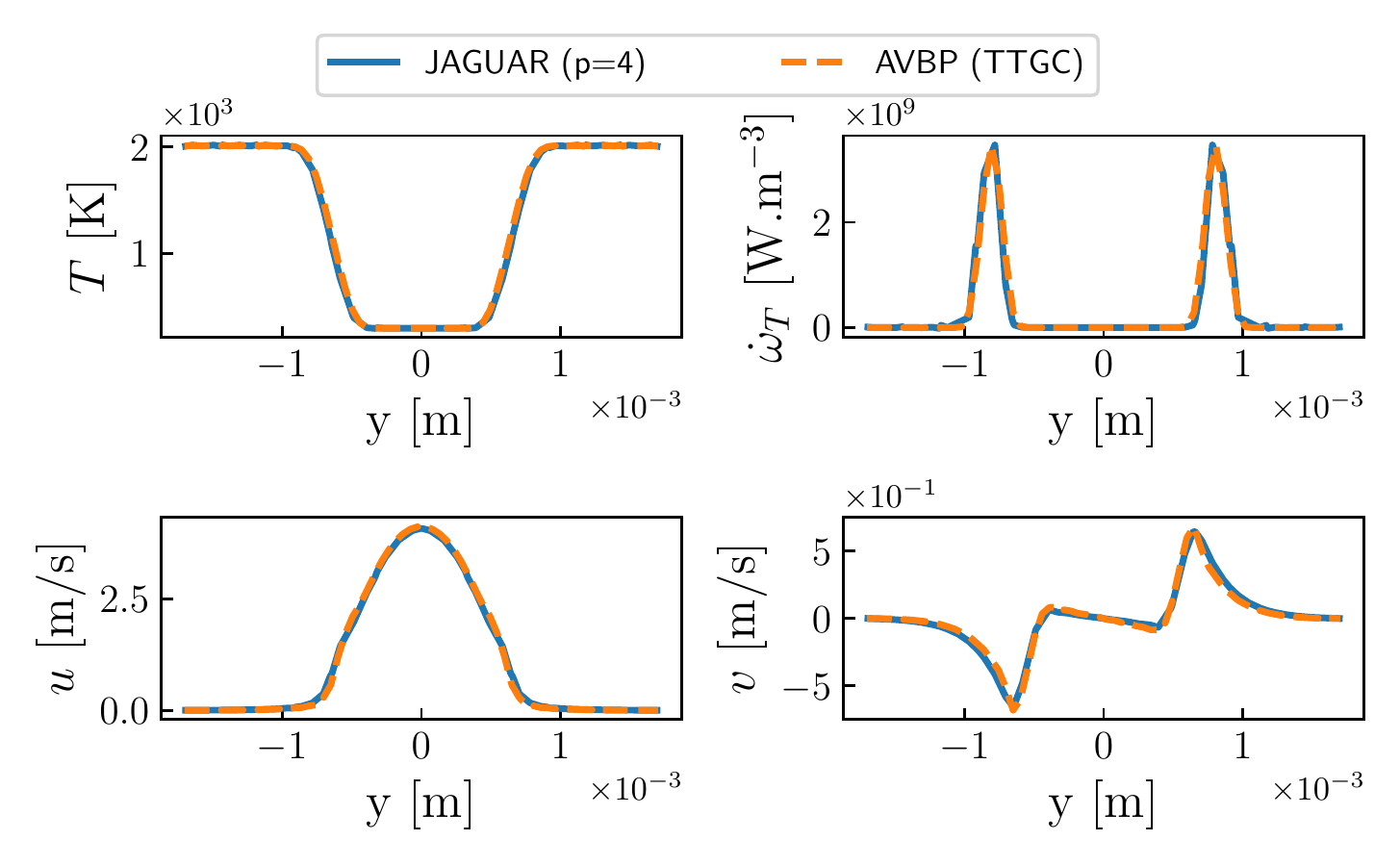}
    \caption{Comparison between JAGUAR and AVBP of temperature, heat release rate, axial and vertical velocity profiles at $x=10.1$ mm along y-axis for the 2D burner case.}
    \label{fig:cut_x10pt1mm_T_hr_u_v}
\end{figure}

\begin{figure}[!h]
    \centering
    \includegraphics[width=\Lfig, height=\Hfig]{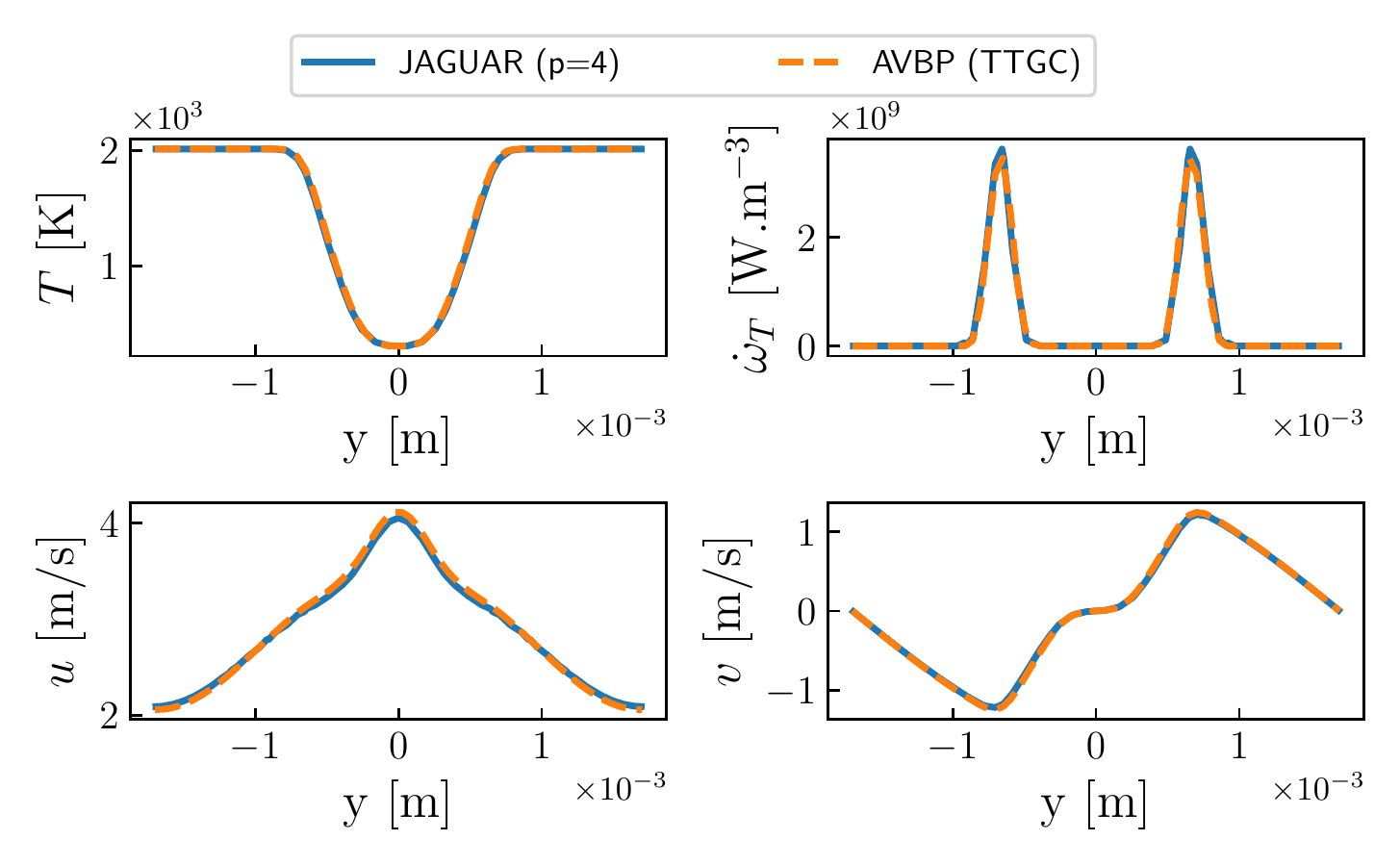}
    \caption{Comparison between JAGUAR and AVBP of temperature, heat release rate, axial and vertical velocity profiles at $x=12$ mm along y-axis for the 2D burner case.}
    \label{fig:cut_x12mm_T_hr_u_v}
\end{figure}

\begin{figure}[!h]
    \centering
    \includegraphics[width=\Lfig, height=\Hfig]{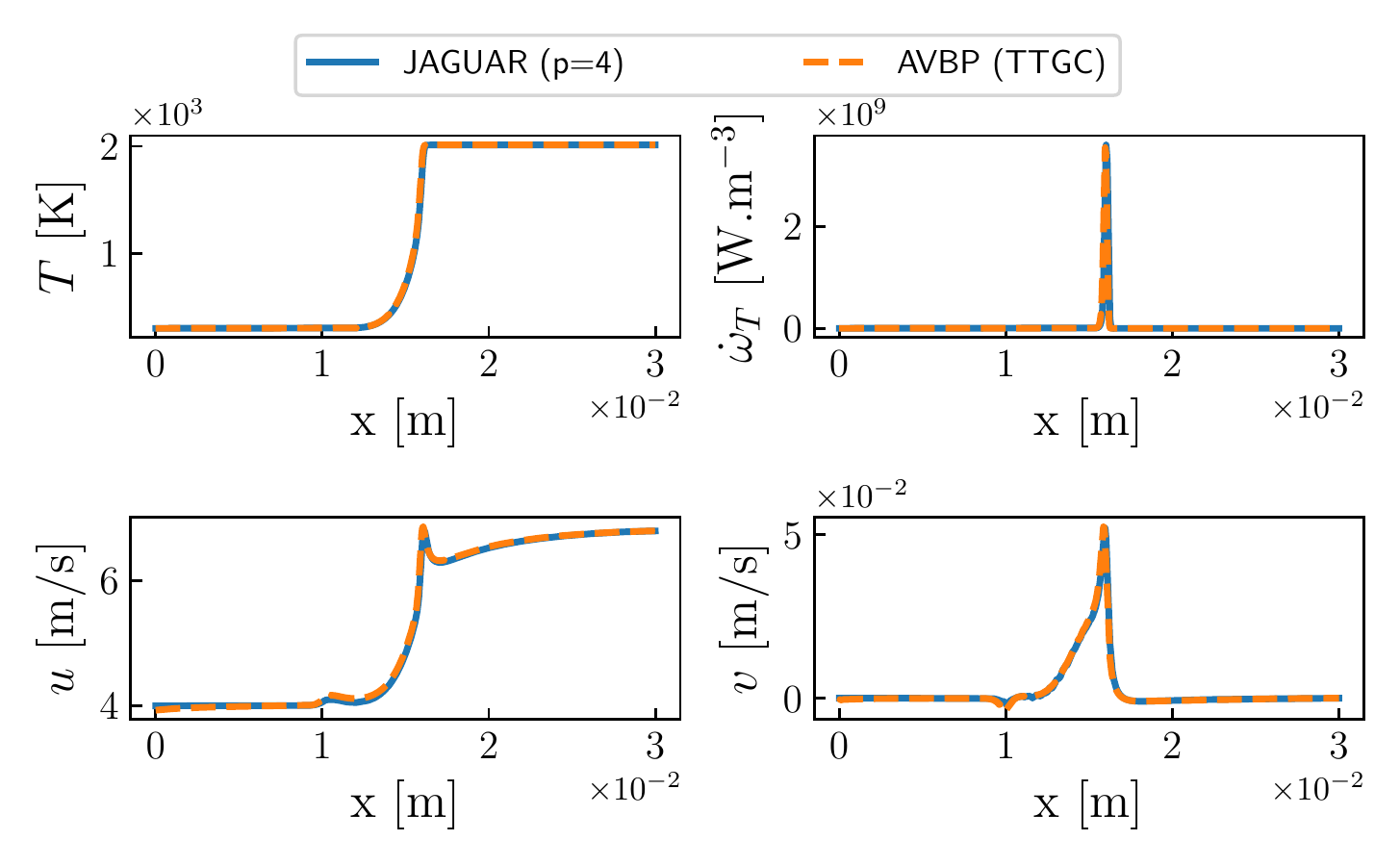}
    \caption{Comparison between JAGUAR and AVBP of temperature, heat release rate, axial and vertical velocity profiles at $y=0$ mm along x-axis for the 2D burner case.}
    \label{fig:cut_y0mm_T_hr_u_v}
\end{figure}

\noindent Figure~\ref{fig:compa_hr_top_JAG_bot_AVBP} shows the 2D heat release rate fields for both JAGUAR and AVBP simulations. They both reproduce the same flame structure. Vertical profiles respectively at $x=10.1$ mm (close to injector outlet) and at $x=12$ mm in Figures~\ref{fig:cut_x10pt1mm_T_hr_u_v} and~\ref{fig:cut_x12mm_T_hr_u_v} are also merely identical between JAGUAR and AVBP. Moreover, as illustrated in Figure~\ref{fig:cut_y0mm_T_hr_u_v}, horizontal profiles along the centerline $y=0$ are also similar with a flame tip at almost the same location: $x_{tip}=15.97$ mm for AVBP and $x_{tip}=16.02$ mm for JAGUAR. Finally, Table~\ref{tab:compa_Tb_maxhr_ref_DNS} shows that both codes predict the same burning temperature $T^{b}$ and almost the same maximum of heat release in the whole domain. CANTERA results obtained for the equivalent 1D flame are also added as reference values.

\begin{table}[!h]
    \centering
    \begin{tabular}{|c|c|c|}
        \hline
       Code & $T^{b}$ [K]  & $\dot{\omega}_{T}^{max}\ \mathrm{[W.m^{-3}]}$ \\
       \hline
       JAGUAR & 2012 & $3.69\times 10^{9}$ \\
       \hline
       AVBP & 2012 & $3.60\times 10^{9}$ \\
       \hline
       CANTERA & 2010 & $3.71\times 10^{9}$ \\
       \hline
    \end{tabular}
    \caption{Comparison between JAGUAR and AVBP on burning flame temperatures $T^{b}$ and maximum heat release rate $\dot{\omega}_{T}^{max}$ obtained in a 2D burner case using the CH4/Air-2S-BFER scheme at $\phi=0.8$. CANTERA values obtained on a 1D flame in the same conditions are also shown as reference.}
    \label{tab:compa_Tb_maxhr_ref_DNS}
\end{table}

\noindent About the computational cost in that case, JAGUAR has an efficiency (cost per iteration and per DOFs) around $7.5\ \mathrm{\mu s/ite/DOFs}$ whereas AVBP using TTGC scheme is around $6.0\ \mathrm{\mu s/ite/DOFs}$. This is quite encouraging since JAGUAR is a very recent code where no real optimization was done compared to AVBP which has more than twenty years of experience in computational fluid dynamics. 

%% file: Conclusion.tex
\section{Conclusion and perspectives\label{sec:conclusion}}
The implementation of combustion source terms, the transport of a multispecies gas and the treatment of boundary conditions in reacting flows have been done within a SD context. Classical SD algorithm which interpolates conservative variables from SP to FP was found unstable in the case of a multispecies gas only. To overcome this issue, it was shown that temperature and pressure must be computed at SP first and then interpolated on FP.

Laminar one and two-dimensional flames have been computed with the SD code JAGUAR and compared to reference combustion codes CANTERA and AVBP. Firstly, one-dimensional flames using either simple two-reactions chemistry or ARC for methane were well computed by JAGUAR which also captured the expected pressure drop across the flame front. In terms of flame speeds, JAGUAR results are slightly better than AVBP and some improvements are seen as the polynomial order increases. Finally, a two-dimensional burner case was simulated to validate wall and symmetry boundary conditions implemented in JAGUAR for a reactive case. The results were very satisfying since JAGUAR and AVBP gave merely the same flame structure. Moreover, the computational cost is only a bit higher than the AVBP one but a lot of optimization can still be done associated with SD features such as $p$-refinement that can considerably reduced this cost.   
To conclude, for all the cases studied here, JAGUAR results are at least as good as AVBP and with almost the same computational cost. It shows that the SD method is able to simulate simple combustion applications such as other classical numerical methods. JAGUAR calculations were done using a constant polynomial order $p$ but future work will focus on the development of $p$-refinement techniques when doing combustion applications with the SD method.

The present work is a starting point for developing the SD method on combustion applications which starts from doing simple laminar test cases and here using a constant polynomial order $p$. Future work will be focused on the development of two main topics:
\begin{itemize}
    \item[$\bullet$] $p$-refinement techniques when doing combustion applications with the SD method.
    \item[$\bullet$] Show the capability of the SD approach to do 3D turbulent combustion test cases.
\end{itemize}

%% file: Appendix_tangential_vectors.tex
\appendix
\section{Tangential vectors for a given normal vector}
\label{appendix:tangential_vectors}
\noindent Given an unit normal vector $\mathbf{n}=\left(n_{x},n_{y},n_{z}\right)^{\mathrm{T}}$ at a boundary FP, two unit vectors can be defined to form a local orthonormal basis at the boundary FP:
\begin{itemize}
    \item[$\bullet$] If $|n_{z}|< 0.7$:
    \begin{equation}
        \mathbf{t_{1}} = \frac{1}{\sqrt{n_{x}^{2}+n_{y}^{2}}}\left(
        \begin{array}{c}
             n_{y}  \\
             -n_{x} \\
             0
        \end{array}
        \right)\hspace{0.25 cm}\text{and}\hspace{0.25 cm} \mathbf{t_{2}} = \frac{1}{\sqrt{n_{x}^{2}+n_{y}^{2}}}\left(
        \begin{array}{c}
             -n_{x}n_{z}  \\
             -n_{y}n_{z} \\
             n_{x}^{2}+n_{y}^{2}
        \end{array}
        \right)
    \end{equation}
    \item[$\bullet$] Else:
    \begin{equation}
        \mathbf{t_{1}} = \frac{1}{\sqrt{n_{y}^{2}+n_{z}^{2}}}\left(
        \begin{array}{c}
             0  \\
             -n_{z} \\
             n_{y}
        \end{array}
        \right)\hspace{0.25 cm}\text{and}\hspace{0.25 cm} \mathbf{t_{2}} = \frac{1}{\sqrt{n_{y}^{2}+n_{z}^{2}}}\left(
        \begin{array}{c}
             n_{y}^{2}+n_{z}^{2}  \\
             -n_{x}n_{y} \\
             -n_{x}n_{z}
        \end{array}
        \right)
    \end{equation}
\end{itemize}

%% file: Appendix_system_to_solve_nscbc_inlet.tex
\section{System of equations for a NSCBC inlet imposing velocities, temperature and species mass fractions}
\label{appendix:T_eqn_inlet}
\noindent The objective is to obtain time-derivatives expressions of the primitive variables $\mathbf{Q}=\left(\rho,u,v,w,P,Y_{1},\hdots,Y_{N_{s}}\right)^{\mathrm{T}}$ as a function of $\boldsymbol{\mathcal{N}}$ and $\boldsymbol{\mathcal{S}}$. To do so, Eq.~(\ref{eq:characteristic_equation_fievet}) is multiplied by $P_{\mathbf{Q}}^{-1}$ whose expression is recalled here:
\begin{align}
    P_{\mathbf{Q}}^{-1} = \left[
    \begin{array}{cccccc}
       n_{x} & n_{y} & n_{z} & \rho/\left(\sqrt{2}c\right) & \rho/\left(\sqrt{2}c\right) & O_{1,N_{s}} \\
        0 & -n_{z} & n_{y} & n_{x}/\sqrt{2} &-n_{x}/\sqrt{2} & O_{1,N_{s}} \\
        n_{z} & 0 & -n_{x} & n_{y}/\sqrt{2} & -n_{y}/\sqrt{2} & O_{1,N_{s}} \\
        -n_{y} & n_{x} & 0 & n_{z}/\sqrt{2} & -n_{z}/\sqrt{2} & O_{1,N_{s}}\\
        0 & 0 & 0 & \rho c/\sqrt{2} & \rho c/\sqrt{2} & O_{1,N_{s}} \\
        O_{N_{s},1} & O_{N_{s},1} & O_{N_{s},1} & O_{N_{s},1} & O_{N_{s},1} & I_{N_{s},N_{s}}
    \end{array}
    \right]
\end{align}
where $O_{m,n}$ is the zero matrix of dimension $m\times n$ and $I_{N_{s},N_{s}}$ is the identity matrix of size $N_{s}$. This multiplication gives:
\begin{align}
    |J|\frac{\partial \rho}{\partial t} &= -n_{x}\left(\mathcal{N}_{1}^{*}+\mathcal{S}_{1}\right)-n_{y}\left(\mathcal{N}_{2}^{*}+\mathcal{S}_{2}\right)-n_{z}\left(\mathcal{N}_{3}^{*}+\mathcal{S}_{3}\right)-\frac{\rho}{\sqrt{2}c}\left(\mathcal{N}_{+}^{*}+\mathcal{S}_{+}+\mathcal{N}_{-}+\mathcal{S}_{-}\right)\\
    |J|\frac{\partial u}{\partial t} &= n_{z}\left(\mathcal{N}_{2}^{*}+\mathcal{S}_{2}\right)-n_{y}\left(\mathcal{N}_{3}^{*}+\mathcal{S}_{3}\right) - \frac{n_{x}}{\sqrt{2}}\left(\mathcal{N}_{+}^{*}+\mathcal{S}_{+}-\mathcal{N}_{-}-\mathcal{S}_{-}\right) \label{eq:inlet_dudt_appendix}\\
    |J|\frac{\partial v}{\partial t} &= -n_{z}\left(\mathcal{N}_{1}^{*}+\mathcal{S}_{1}\right)+n_{x}\left(\mathcal{N}_{3}^{*}+\mathcal{S}_{3}\right) - \frac{n_{y}}{\sqrt{2}}\left(\mathcal{N}_{+}^{*}+\mathcal{S}_{+}-\mathcal{N}_{-}-\mathcal{S}_{-}\right) \label{eq:inlet_dvdt_appendix}\\
    |J|\frac{\partial w}{\partial t} &= n_{y}\left(\mathcal{N}_{1}^{*}+\mathcal{S}_{1}\right)-n_{x}\left(\mathcal{N}_{2}^{*}+\mathcal{S}_{2}\right) - \frac{n_{z}}{\sqrt{2}}\left(\mathcal{N}_{+}^{*}+\mathcal{S}_{+}-\mathcal{N}_{-}-\mathcal{S}_{-}\right) \label{eq:inlet_dwdt_appendix}\\
    |J|\frac{\partial P}{\partial t} &=-\frac{\rho c}{\sqrt{2}}\left(\mathcal{N}_{+}^{*}+\mathcal{S}_{+}+\mathcal{N}_{-}+\mathcal{S}_{-}\right) \\
    |J|\frac{\partial Y_{k}}{\partial t} &=-\left(\mathcal{N}_{5+k}^{*}+\mathcal{S}_{5+k}\right)\hspace{0.20 cm}\text{for}\hspace{0.20 cm} k=1,N_{s}\label{eq:inlet_dYkdt_appendix}
\end{align}
Then, time-derivatives of $u$, $v$ and $w$ are given in Eqs.~(\ref{eq:inlet_dudt_appendix}-\ref{eq:inlet_dwdt_appendix}) and those of $Y_{k}$ in Eq.~(\ref{eq:inlet_dYkdt_appendix}). It remains to find an expression for $|J|\left(\partial T/\partial t\right)$. Thanks to Eq.~(\ref{eq:perfect_gas_law}) the differential of $T$ is given by:
\begin{equation}
\begin{aligned}
     \partial T &= \frac{T}{P}\partial P - \frac{T}{\rho}\partial \rho - TW\partial \left(\frac{1}{W}\right) = \frac{T}{P}\partial P - \frac{T}{\rho}\partial \rho - TW\sum\limits_{k=1}^{N_{s}}{\frac{\partial Y_{k}}{W_{k}}} \\
     \Rightarrow
     |J|\frac{\partial T}{\partial t} &= \frac{T}{P}|J|\frac{\partial P}{\partial t} - \frac{T}{\rho}|J|\frac{\partial \rho}{\partial t} - TW\sum\limits_{k=1}^{N_{s}}{\frac{1}{W_{k}}|J|\frac{\partial Y_{k}}{\partial t}} \\
     \Rightarrow
     |J|\frac{\partial T}{\partial t} &= -\frac{T\rho c}{\sqrt{2}P}\left(\mathcal{N}_{+}^{*}+\mathcal{S}_{+}+\mathcal{N}_{-}+\mathcal{S}_{-}\right) \\
     &+ \frac{T}{\rho}\left[n_{x}\left(\mathcal{N}_{1}^{*}+\mathcal{S}_{1}\right)+n_{y}\left(\mathcal{N}_{2}^{*}+\mathcal{S}_{2}\right)+n_{z}\left(\mathcal{N}_{3}^{*}+\mathcal{S}_{3}\right)+\frac{\rho}{\sqrt{2}c}\left(\mathcal{N}_{+}^{*}+\mathcal{S}_{+}+\mathcal{N}_{-}+\mathcal{S}_{-}\right)\right] \\
     &+ TW\sum\limits_{k=1}^{N_{s}}{\frac{\mathcal{N}_{5+k}^{*}+\mathcal{S}_{5+k}}{W_{k}}} \\
     \Rightarrow
     |J|\frac{\partial T}{\partial t} &= \frac{T}{\rho}\left[n_{x}\left(\mathcal{N}_{1}^{*}+\mathcal{S}_{1}\right)+n_{y}\left(\mathcal{N}_{2}^{*}+\mathcal{S}_{2}\right)+n_{z}\left(\mathcal{N}_{3}^{*}+\mathcal{S}_{3}\right)\right]-\frac{T}{\sqrt{2}}\underbrace{\left(\frac{\rho c}{P}-\frac{1}{c}\right)}_{\frac{\gamma-1}{c}}\left(\mathcal{N}_{+}^{*}+\mathcal{S}_{+}+\mathcal{N}_{-}+\mathcal{S}_{-}\right) \\
     &+ TW\sum\limits_{k=1}^{N_{s}}{\frac{\mathcal{N}_{5+k}^{*}+\mathcal{S}_{5+k}}{W_{k}}} \\
     \Rightarrow
     |J|\frac{\partial T}{\partial t} &= \frac{T}{\rho}\left[n_{x}\left(\mathcal{N}_{1}^{*}+\mathcal{S}_{1}\right)+n_{y}\left(\mathcal{N}_{2}^{*}+\mathcal{S}_{2}\right)+n_{z}\left(\mathcal{N}_{3}^{*}+\mathcal{S}_{3}\right)\right]-\frac{T\left(\gamma -1\right)}{\sqrt{2}c}\left(\mathcal{N}_{+}^{*}+\mathcal{S}_{+}+\mathcal{N}_{-}+\mathcal{S}_{-}\right) \\
     &+ TW\sum\limits_{k=1}^{N_{s}}{\frac{\mathcal{N}_{5+k}^{*}+\mathcal{S}_{5+k}}{W_{k}}}
     \label{eq:compute_dT_dt_1}
\end{aligned}
\end{equation}

%% file: Appendix_get_dQdU.tex
\section{Transformation matrices from conservative (respectively primitive) to primitive (respectively conservative) variables for a multispecies thermally perfect gas}
\label{appendix:get_dQdU}
\noindent Let's denote by $\mathbf{U}=\left(\rho,\rho u,\rho v,\rho w,\rho E,\rho Y_{1},\hdots,\rho Y_{N_{s}}\right)^{\mathrm{T}}$ the vector of conservative variables and $\mathbf{Q}=\left(\rho,u,v,w,P,Y_{1},\hdots,Y_{N_{s}}\right)^{\mathrm{T}}$ the vector of primitive variables with density as first variable and pressure as fifth variable. For a multispecies thermally perfact gas, $\rho E$ is the sum of sensible and kinetic energies per unit of volume:
\begin{align}
    \rho E = \rho e_{s} + \rho\frac{||\mathbf{u}||^{2}}{2}
    \label{eq:def_rhoE_multi_thermally}
\end{align}
where $e_{s}$ is related to the sensible enthalpy $h_{s}$ through Eq.~(\ref{eq:def_es_multi_thermally}) \cite{poinsot2005theoretical}:
\begin{align}
    e_{s} = h_{s} - \frac{P}{\rho} = \sum\limits_{k=1}^{N_{s}}{Y_{k}h_{sk}} - \frac{P}{\rho}
    \label{eq:def_es_multi_thermally}
\end{align}
with $h_{sk}\equiv h_{sk}\left(T\right)=\int_{T_{0}}^{T}{C_{pk}\left(T'\right)dT'}$, $C_{pk}$ being the heat capacity at constant pressure of species $k$ and $T_{0}$ a reference temperature. Moreover, the multispecies gas is also assumed to behave as and ideal gas so that Eq.~(\ref{eq:perfect_gas_law}) holds true. The objective is to compute the transformation matrix from conservative to primitive variables for this kind of gas defined as:
\begin{align}
    \frac{\partial \mathbf{Q}}{\partial \mathbf{U}} \equiv \frac{\partial \left(\rho,u,v,w,P,Y_{1},\hdots,Y_{N_{s}}\right)}{\partial \left(\rho,\rho u,\rho v,\rho w,\rho E,\rho Y_{1},\hdots,\rho Y_{N_{s}}\right)}
\end{align}
Firstly, the transformation matrix from primitive to conservative variables defined as:
\begin{align}
    \frac{\partial \mathbf{U}}{\partial \mathbf{Q}} \equiv \frac{\partial \left(\rho,\rho u,\rho v,\rho w,\rho E,\rho Y_{1},\hdots,\rho Y_{N_{s}}\right)}{\partial \left(\rho,u,v,w,P,Y_{1},\hdots,Y_{N_{s}}\right)}
\end{align}
will be computed and then it will be inverted to get $\left(\partial \mathbf{Q}/\partial \mathbf{U}\right)$. Compare to a calorically perfect gas, the four first lines of $\left(\partial \mathbf{U}/\partial \mathbf{Q}\right)$ are exactly the same: the differences lie in the fifth line because the expression of $\rho E$ is different for a thermally perfect gas.
\subsection{Computation of $\left(\partial \rho E/\partial \rho\right)$ for $\left(\partial \mathbf{U}/\partial \mathbf{Q}\right)$ matrix}
\noindent Starting from Eqs.~(\ref{eq:def_rhoE_multi_thermally}-\ref{eq:def_es_multi_thermally}), $\left(\partial \rho E/\partial \rho\right)$ is given by:
\begin{align}
     \frac{\partial \rho E}{\partial \rho} = e_{s} + \rho\frac{\partial e_{s}}{\partial \rho} + \frac{||\mathbf{u}||^{2}}{2} = e_{s} + \rho\left(\sum\limits_{k=1}^{N_{s}}{Y_{k}\frac{\partial h_{sk}}{\partial \rho}} + \frac{P}{\rho^{2}}\right) + \frac{||\mathbf{u}||^{2}}{2} = e_{s} + \rho\left(\sum\limits_{k=1}^{N_{s}}{Y_{k}\frac{\partial h_{sk}}{\partial T}\frac{\partial T}{\partial \rho}} + \frac{P}{\rho^{2}}\right) + \frac{||\mathbf{u}||^{2}}{2}
     \label{eq:drhoEdrho_1}
\end{align}
where by definition $\left(\partial h_{sk}/\partial T\right)=C_{pk}$ and $\left(\partial T/\partial \rho\right)=-T/\rho$ according to Eq.~(\ref{eq:perfect_gas_law}). Consequently, Eq.~(\ref{eq:drhoEdrho_1}) becomes:
\begin{align}
    \frac{\partial \rho E}{\partial \rho} = \underbrace{e_{s} + \frac{P}{\rho}}_{h_{s}} -T\underbrace{\sum\limits_{k=1}^{N_{s}}{Y_{k}C_{pk}}}_{C_{p}} + \frac{||\mathbf{u}||^{2}}{2} = h_{s} -TC_{p} + \frac{||\mathbf{u}||^{2}}{2}
\end{align}

\subsection{Computation of $\left(\partial \rho E/\partial P\right)$ for $\left(\partial \mathbf{U}/\partial \mathbf{Q}\right)$ matrix}
\noindent Starting again from Eqs.~(\ref{eq:def_rhoE_multi_thermally}-\ref{eq:def_es_multi_thermally}), $\left(\partial \rho E/\partial P\right)$ is given by:
\begin{align}
    \frac{\partial \rho E}{\partial P} = \rho\frac{\partial e_{s}}{\partial P} = \rho\left(\sum\limits_{k=1}^{N_{s}}{Y_{k}\frac{\partial h_{sk}}{\partial P}} - \frac{1}{\rho}\right) = \rho\left(\sum\limits_{k=1}^{N_{s}}{Y_{k}\underbrace{\frac{\partial h_{sk}}{\partial T}}_{C_{pk}}\underbrace{\frac{\partial T}{\partial P}}_{\frac{W}{\rho \overline{R}}}} - \frac{1}{\rho}\right) = \frac{W}{\overline{R}}C_{p}-1 \label{eq:drhoEdP_1}
\end{align}
and since $\overline{R}/W=C_{p}-C_{v}$, Eq.~(\ref{eq:drhoEdP_1}) becomes:
\begin{align}
    \frac{\partial \rho E}{\partial P} = \frac{C_{p}}{C_{p}-C_{v}} - 1 = \frac{1}{1-\frac{1}{\gamma}} - 1 = \frac{\gamma - \gamma + 1}{\gamma-1} = \frac{1}{\gamma-1}
\end{align}
which is the same results for $\left(\partial \rho E/\partial P\right)$ when the gas is calorically perfect (constant $C_{p}$ and $C_{v}$).

\subsection{Computation of $\left(\partial \rho E/\partial Y_{k}\right)$ for $\left(\partial \mathbf{U}/\partial \mathbf{Q}\right)$ matrix}
\noindent Starting again from Eqs.~(\ref{eq:def_rhoE_multi_thermally}-\ref{eq:def_es_multi_thermally}), $\left(\partial \rho E/\partial Y_{k}\right)$ is given by:
\begin{align}
    \frac{\partial \rho E}{\partial Y_{k}} = \rho\frac{\partial e_{s}}{\partial Y_{k}} = \rho\left(h_{sk} + \sum\limits_{k'=1}^{N_{s}}{Y_{k'}\frac{\partial h_{sk'}}{\partial Y_{k}}}\right) = \rho\frac{\partial e_{s}}{\partial Y_{k}} = \rho\left(h_{sk} + \sum\limits_{k'=1}^{N_{s}}{Y_{k'}\frac{\partial h_{sk'}}{\partial T}\frac{\partial T}{\partial Y_{k}}}\right) \label{eq:drhoEdYk_1}
\end{align}
and since $\left(\partial T/\partial Y_{k}\right)=\left[P/\left(\rho \overline{R}\right)\right]\left(\partial W/\partial Y_{k}\right)=-\left[PW^{2}/\left(\rho \overline{R}\right)\right]\left(\partial \left(1/W\right)/\partial Y_{k}\right)=-TW/W_{k}$, Eq.~(\ref{eq:drhoEdYk_1}) becomes:
\begin{align}
    \frac{\partial \rho E}{\partial Y_{k}} = \rho\left(h_{sk} - \frac{TW}{W_{k}}\sum\limits_{k'=1}^{N_{s}}{Y_{k'}C_{pk'}}\right) = \rho\left(h_{sk} - TC_{p}\frac{W}{W_{k}}\right)
\end{align}
Consequently, $\left(\partial \mathbf{U}/\partial \mathbf{Q}\right)$ for a multispecies thermally perfect gas is given by Eq.~(\ref{eq:dUdQ_matrix}):
\begin{equation}
    \frac{\partial \mathbf{U}}{\partial \mathbf{Q}} = \left[
    \begin{array}{cccccccc}
        1 & 0 & 0 & 0 & 0 & O_{1,N_{s}} & & \\
        u & \rho & 0 & 0 & 0 & O_{1,N_{s}} & &\\
        v & 0 & \rho & 0 & 0& O_{1,N_{s}} & &\\
        w & 0 & 0 & \rho & 0 & O_{1,N_{s}} & &\\
        h_{s}-TC_{p}+\frac{||\mathbf{u}||^{2}}{2} & \rho u & \rho v & \rho w & \frac{1}{\gamma-1} & \frac{\partial \rho E}{\partial Y_{1}} & \hdots & \frac{\partial \rho E}{\partial Y_{N_{s}}}\\
        Y_{1} & O_{N_{s},1} & O_{N_{s},1} & O_{N_{s},1} & O_{N_{s},1} &   & & \\
        \vdots &  &  &  &  &  & \rho I_{N_{s},N_{s}}& \\
        Y_{N_{s}} &  &  &  &  &   & &
    \end{array}
    \right]
    \label{eq:dUdQ_matrix}
\end{equation}
Finally, the inverse matrix of $\left(\partial \mathbf{U}/\partial \mathbf{Q}\right)$ is $\left(\partial \mathbf{Q}/\partial \mathbf{U}\right)$ whose expression is given by Eq.~(\ref{eq:dQdU_matrix}).